\documentclass[mlmain]{jmlr}
\jmlrpmlr

\usepackage{booktabs}
\usepackage{array}
\usepackage{float}

\newcommand{\tobs}{\theta_{\mathrm{obs}}}
\newcommand{\tfloor}{\theta_{\mathrm{floor}}}
\newcommand{\PR}{\mathrm{PR}}
\jmlrvolume{}
\firstpageno{1}
\editors{List of editors' names}
\hypersetup{pdftitle={Structure is in the zoom: probing neural symmetry through dimensionality scaling}, pdfauthor={Adil Amin}}

\jmlryear{2026}
\jmlrworkshop{Symmetry and Geometry in Neural Representations}

\title[Structure is in the zoom]{Structure is in the zoom: probing neural symmetry\\through dimensionality scaling}

\author{\Name{Adil Amin} \Email{adil@zehenlabs.com}\\
\addr ZEHEN Labs, Milwaukee, WI, USA}

\begin{document}

\maketitle

\begin{abstract}\tolerance=400
Scaling exponents of neural dimensionality are read as properties of the code, and used to compare representations and benchmark models against cortex. Yet one recording of 11{,}311 mouse V1 neurons gives $0.25$, $0.30$ or $0.35$, depending only on how it is subsampled.
We show that such an exponent measures the population and the probe together. We split it exactly into a floor, which random subsets of the same sizes give, and a shift, which a declared axis of stimulus classes adds. The usual random-subsample exponent is almost all floor. Stimulus direction lifts the exponent in every grating recording by reading both symmetry sectors of the code: orientation, which repeats every half turn, and direction, which does not.
The lift lives in trial-to-trial fluctuations that differ between directions, while the class means, which linear decoders read, pull it down. Two measured terms, how much the classes' fluctuations overlap and how the means sit against them, account for the shift. They give its sign where it changes: in Allen Neuropixels recordings, whose class means mostly raise the exponent, and in macaque V1. In an XY lattice whose classes stay on one orbit, temperature flips the sign as the two terms say, and in GPT-2 the shift ranks the indirect-object circuit's name-mover and backup heads.
Reported with its path, its floor and its shift, an exponent says how much structure a probe finds and what carries it. Our open tool, \texttt{rung}, reports all three for any labelled neural recording or language-model layer: \url{https://github.com/adilamin89/structure-in-the-zoom}.
\end{abstract}

\section{Introduction}
\label{sec:intro}

A common way to characterize a neural population is to ask how its dimensionality grows with the data used to measure it. The participation ratio, $\PR = (\sum_i\lambda_i)^2/\sum_i\lambda_i^2$ for covariance eigenvalues $\lambda_i$, counts the dimensions of comparable variance the activity occupies \citep{gao2017}. Measured on subsets of growing size, PR grows, and its log-log slope $\tobs$ is reported as the population's scaling exponent. Like the decay exponent of the cross-validated spectrum \citep{stringer2019nature}, it is read as a property of the code. Dimension estimates are used to compare representations, benchmark models against cortex and reason about computation \citep{cunningham2014,chung2018,jazayeri2021,bernardi2020,elmoznino2024,canatar2023}. Every comparison built on such an exponent rests on whether it reflects the code or the measurement.

\begin{figure}[t]
\centering
\includegraphics[width=\linewidth]{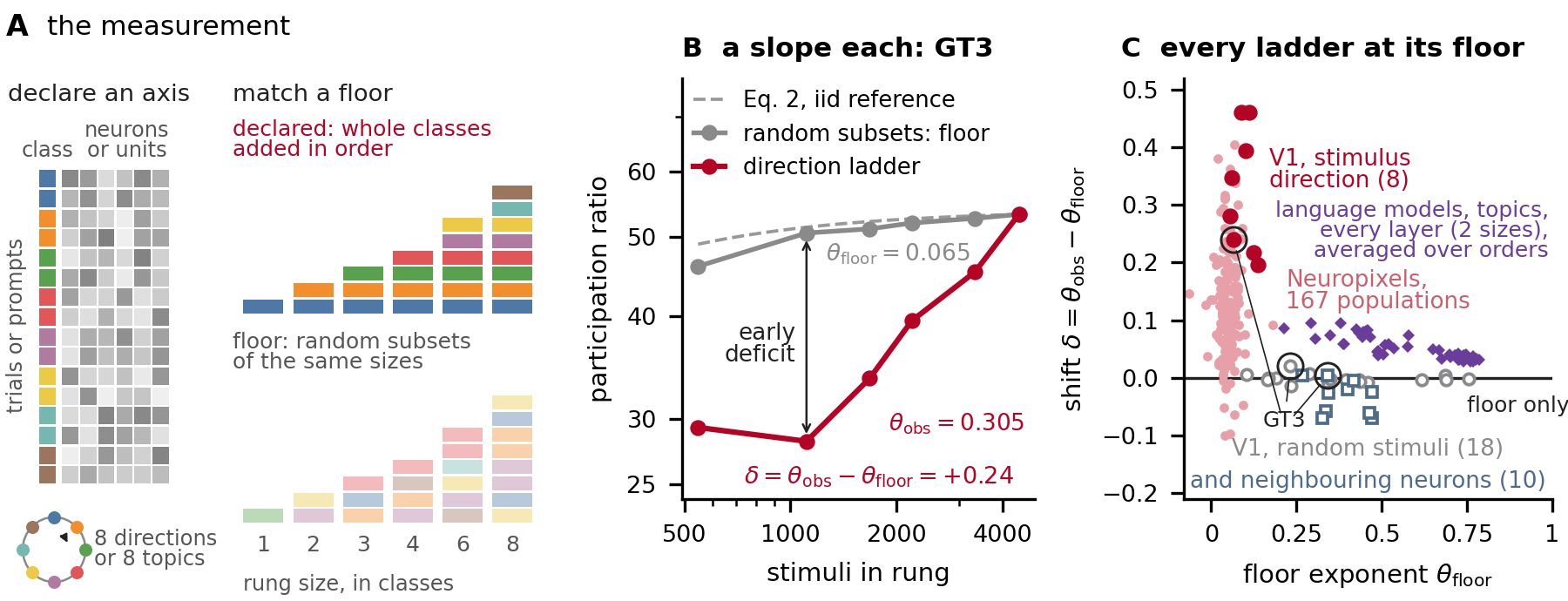}
\caption{One population, many probes. (\textbf{A})~A declared axis adds whole classes in order; its floor draws random subsets of the same sizes. (\textbf{B})~GT3: the floor (gray; Eq.~\ref{eq:null} dashed) and the direction ladder (red), whose slopes split as $\tobs = \tfloor + \delta$. (\textbf{C})~Every ladder's shift against its floor (GT3 ringed). Ladders on the zero line are all floor; above it, directions lift all eight grating recordings and 158 of 167 Neuropixels populations (temporal frequencies pooled), topics every layer of Pythia-160m and 2.8B (order-averaged; Appendix~\ref{app:llm}).}
\label{fig:ladder}
\end{figure}

One recording of 11{,}311 mouse V1 neurons returns three exponents (Figure~\ref{fig:ladder}). Adding random stimuli gives $0.25$, adding stimulus directions in angular order $0.30$, and adding neighbouring neurons $0.35$. Which one is the population's exponent?

Sampling is known to shape the reading. Random subsets trace how dimensionality approaches an asymptote \citep{lehky2014,williamson2016}, and which neurons are sampled changes the spectrum \citep{kang2025} and how dimensionality grows \citep{wang2025}. Temporal correlation and limited sampling can mimic scaling \citep{calvo2026,levina2022}, and recording choices move the coarse-graining exponents of identical dynamics \citep{li2026}. Estimators correct for finite trials and neurons \citep{chun2026,pospisil2025}, and theory asks what sets dimensionality \citep{dahmen2026}. The closest precedents change what is added. \citet{cowley2016} accumulated grating directions in macaque V1: the dimensionality of trial-averaged responses stayed below the chance level of random patterns and changed slope after half the circle. \citet{mazzucato2016} argued that sampling a clustered network's neurons one per cluster changes how dimensionality grows. Neither reads a path through whole classes against random subsets of the same population at the same sizes. That comparison separates the part of a slope a path adds from the part sampling alone gives.

We make the path part of the measurement. Just as a diffraction experiment reads a crystal's order only along the wavevectors it probes, a subsampling path reads a population's structure only along the variable it follows. We split the exponent exactly into a floor, what label-blind random subsets of the same sizes return, and a shift $\delta$, what a declared axis adds. With the floor subtracted, the declared axis becomes an experimental variable and the shift its reading. When the classes form a group orbit, as directions on a circle do, the shift reads the covariance geometry along the orbit: which symmetry sectors the class means occupy, and how the fluctuations change across it. Split this way, the random-stimulus and neighbouring-neuron exponents of Figure~\ref{fig:ladder} are floors, and only the path along direction rises above its own. The population is the same in all three; the structure is in the zoom.

We show that the usual random-subsample exponent is almost all floor (Section~\ref{sec:results}). Stimulus direction lifts the exponent above its floor in every grating recording, reading both symmetry sectors of the code, orientation and direction (Sections~\ref{sec:results} and~\ref{sec:symmetry}). The lift lives in trial-to-trial fluctuations that differ between directions; the class means, the part a linear decoder (a linear probe, in machine-learning terms) reads, pull the exponent down (Section~\ref{sec:sign}). Two measured terms account for the shift: how much the classes' fluctuations overlap, and how the means sit against them (Section~\ref{sec:account}). They give its sign where it changes, in a second laboratory and in macaque V1 (Section~\ref{sec:allen}). Where the answer is known, in lattice models, a rotation-invariant network and a language-model circuit, the shift reads it back (Section~\ref{sec:observable}). Beside a decoder, which says whether a label can be read out, the shift says how much of an exponent the label accounts for and what sets that share.

We release the measurement as an open tool, \texttt{rung} \citep{rung2026}, for any samples-by-features array with row labels, including every layer or attention head of a language model. Appendix~\ref{app:worked} works one recording through it; Table~\ref{tab:roadmap} maps each appendix to its claim.

\section{The measurement: a declared axis against its floor}
\label{sec:decomp}

An exponent is an instrument reading, and the instrument is the ladder: subsets of growing size $N_k$, of presentations or of units, with PR measured on each (Figure~\ref{fig:ladder}A). Two ladders climb through the same recording. One ignores the labels and draws each rung at random; its slope is the floor. The other follows a declared axis, a variable that sorts the samples into classes entered in a fixed order, and each of its rungs adds whole classes. Where classes occupy different subspaces, the declared ladder starts low, because one class spans few dimensions, and climbs as each class brings new ones; the floor climbs only as sampling fills the spectrum. The shift is the difference of the two slopes (Figure~\ref{fig:ladder}B). When it is positive, the axis lifts the exponent above its floor.

Write $y_k = \log \PR^{(k)}$ for rung $k$ of the declared ladder (each subset centered on its mean; Appendix~\ref{app:methods}), $f_k$ for the mean log PR of random subsets of the same size (ten random ladders) and $x_k = \log N_k$. The shift $\delta$ is the least-squares slope of $y - f$ on $x$. A least-squares slope is linear in its response, so
\begin{equation}
  \tobs = \tfloor + \delta,
  \label{eq:decomp}
\end{equation}
when the two ladders share rung sizes (Appendix~\ref{app:identity}). The \emph{matched-null floor} $\tfloor$, the slope of $f$, is what the same population, estimator and rung sizes return with labels ignored. Label permutations at fixed rung sizes give the shift's null.

Even iid samples have a floor, because a covariance estimated from $n$ samples cannot show more dimensions than the samples span. Let $S$ be the centered sample covariance of $n$ Gaussian samples. From Wishart trace moments we derive (Appendix~\ref{app:wishart}) that the ratio of expectations over samples, $(\mathbb{E}\operatorname{Tr}S)^2/\mathbb{E}\operatorname{Tr}S^2$, depends on the spectrum only through the population's PR, $d_{\mathrm{eff}} = (\sum\lambda_i)^2/\sum\lambda_i^2$:
\begin{equation}
  \PR_{\mathrm{null}}(n) = \frac{d_{\mathrm{eff}}\,(n-1)}{d_{\mathrm{eff}} + n}.
  \label{eq:null}
\end{equation}
To leading order this is the finite-trial result of \citet[Eq.~16]{mazzucato2016} and \citet[Eq.~1]{landau2026}. For $n \gg 1$, $1/\PR_{\mathrm{null}} \approx 1/n + 1/d_{\mathrm{eff}}$: sample count and dimension combine like resistors in parallel, and PR follows the smaller, the finite-sample crossover of the Marchenko--Pastur law \citep{marchenko1967}. A ladder whose rungs straddle the crossover returns a nonzero slope from iid samples. The shift has a threshold of its own: until each class holds about as many samples as its own fluctuation dimension, the shift can read near zero where structure exists (Appendix~\ref{app:random}).

\section{Three paths through one population}
\label{sec:results}

Of the three exponents of GT3, one mouse's drifting-grating recording, only direction rises far above its floor (Figure~\ref{fig:ladder}C). Random stimuli return $0.253$, nearly all floor (shift $+0.020$), and neighbouring neurons $0.346$, almost all floor ($0.342$). Direction's $0.305$ is a floor of $0.065$ and a shift of $+0.240$. The floors differ because the ladders start in different places. The random ladder starts at 50 presentations, on the crossover of Eq.~\ref{eq:null}, where the floor climbs; the direction ladder starts at one class, 547 presentations, where it is nearly flat.

A random-subsample exponent is almost entirely floor. The imaging data \citep{stringer2021,stringer2019science} hold 9{,}392 to 22{,}565 mouse V1 neurons per recording: eight grating recordings from three mice (GT1--GT3), 390 to 681 presentations per direction class, and ten static-grating or spontaneous ones (Table~\ref{tab:data}). In all 18, random exponents from $0.11$ to $0.75$ track their floors (mean shift $-0.002$), and the random ladders' local slopes fall with $n/d_{\mathrm{eff}}$ in the shape Eq.~\ref{eq:null} predicts (Appendix~\ref{app:random}). Their spread across recordings mostly reflects where each recording's rungs sit against its effective dimension. Move the rungs, and the same population returns another exponent. Neighbouring-neuron ladders, the path of \citet{manley2024} and \citet{wang2025}, stay near their floors (mean shift $-0.033$; Appendix~\ref{app:neuron}), as expected where neighbours' preferences are mixed \citep{ohki2005}.

\paragraph{Direction lifts every grating recording.} Adding the eight direction bins in angular order from $0^\circ$, in rungs of 1, 2, 3, 4, 6 and 8 bins, gives positive shifts from $0.20$ at low contrast to $0.45$ for localized gratings. Each exceeds all 200 label permutations at the same rung sizes ($p = 1/201$; Appendix~\ref{app:orient}). The grouping into directions makes the shift: forty random class orders give positive shifts in all 320 cases, the angular order inside their spread. Orientation classes, which merge opposite directions, give smaller shifts in all eight.

\section{What the direction axis reads: two symmetry sectors}
\label{sec:symmetry}

The direction axis reads two kinds of tuning \citep{hubel1962}. A direction class carries its orientation, shared with the opposite direction, and its own drift direction; harmonics of the class-mean correlation separate the two. Orientation is $\pi$-periodic \citep{benyishai1995}: gratings at $0^\circ$ and $180^\circ$ differ only in drift direction, so the orientation part of the response lives on $\mathbb{RP}^1 = S^1/\mathbb{Z}_2$, the circle with opposite points identified. The correlation $C(\Delta\phi)$ across neurons between two class means depends almost only on their separation $\Delta\phi$ (Appendix~\ref{app:methods}). It is therefore a cosine series, and its harmonics label the irreducible representations of $O(2)$: each coefficient is its sector's share of the normalized class-mean power. Even harmonics are invariant under the antipodal map and encode \emph{orientation} (the \textbf{quadrupole}, $\ell = 2$); odd harmonics change sign and encode \emph{direction} (the \textbf{dipole}, $\ell = 1$).

\begin{figure}[t]
\centering
\includegraphics[width=\linewidth]{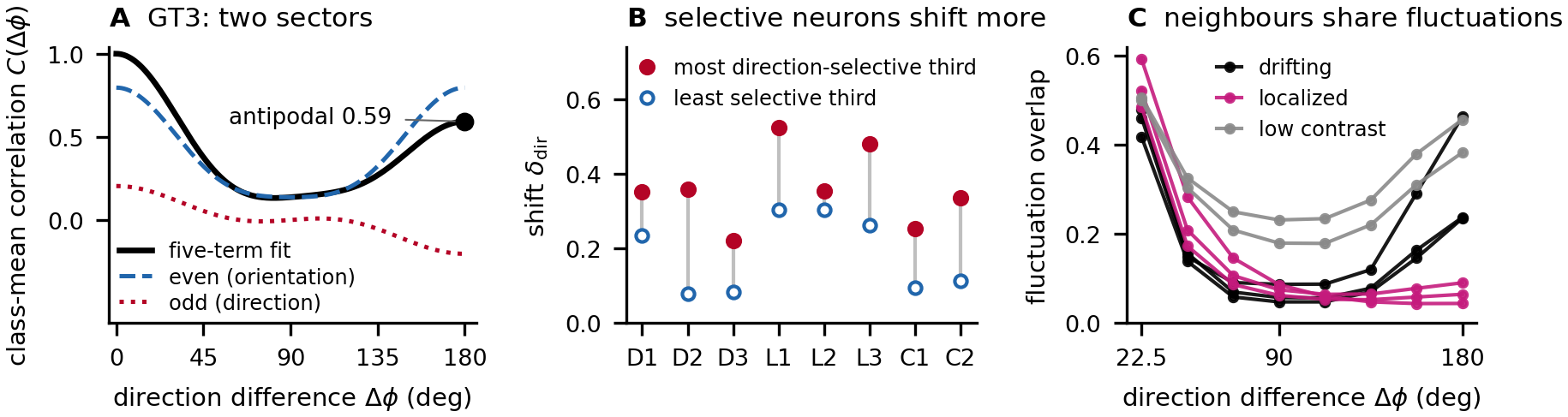}
\caption{What the direction axis reads. (\textbf{A})~GT3's class-mean correlation, with its even (orientation) and odd (direction) sectors: the means are mostly orientation. (\textbf{B})~The most direction-selective third of tuned neurons shifts more than the least in every recording (D drifting, L localized, C low contrast; digit: mouse). (\textbf{C})~Fluctuation overlap of two of sixteen classes against their separation: in full-field recordings opposite directions share more than orthogonal ones.}
\label{fig:multipole}
\end{figure}

The direction axis reads the odd sector where neurons carry it. GT3's class means are mostly orientation (Figure~\ref{fig:multipole}A): their correlation falls from $1$ to $0.14$ at $90^\circ$ and recovers to $0.59$ at $180^\circ$. In every full-field recording (drifting and low-contrast) the quadrupole leads; in the three localized recordings, with more direction-selective neurons ($72$--$78\%$ of tuned neurons against $38$--$61\%$), the dipole leads and the shift is largest. Within every recording the most direction-selective third of tuned neurons shifts more than the least ($0.22$--$0.52$ against $0.08$--$0.30$, also on held-out trials; Figure~\ref{fig:multipole}B). Exactly even class means would return to a correlation of 1 at $180^\circ$. Direction-selective neurons make opposite directions differ: through their means in drifting recordings, their fluctuations in localized ones (Appendix~\ref{app:methods}).

\section{What carries the shift: fluctuations that differ between directions}
\label{sec:sign}

\begin{figure}[t]
\centering
\includegraphics[width=\linewidth]{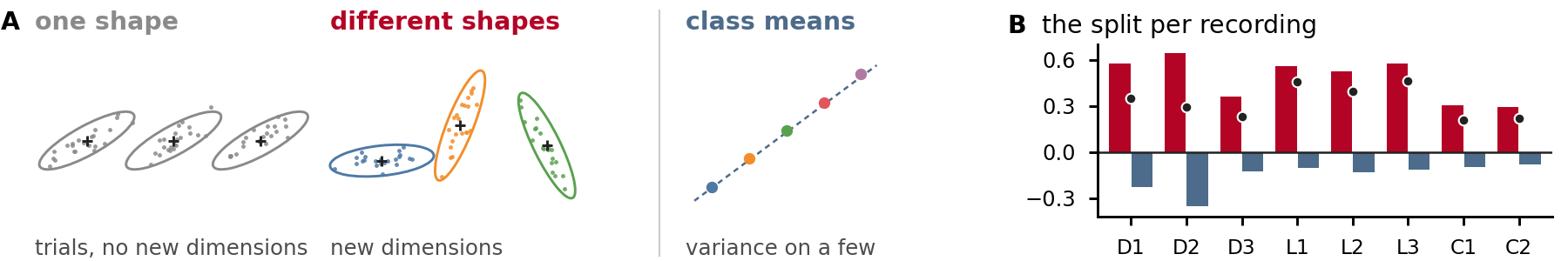}
\caption{What carries the shift. (\textbf{A})~Fluctuations as clouds around the means. (\textbf{B})~The split (Table~\ref{tab:split}): within-class part (red), class-mean part (slate), shift (dots).}
\label{fig:mechanism}
\end{figure}

The trial-to-trial fluctuations around the class means carry the lift, and trial-averaged accumulations \citep{cowley2016} discard them. Picture each class's fluctuations as a cloud around its mean (Figure~\ref{fig:mechanism}A). The declared ladder adds whole clouds, while the floor samples every cloud from its first rung. Clouds of one shape add trials and no new dimensions (left); clouds stretched different ways each add weight along dimensions the others held weakly (middle), so the declared ladder climbs faster than its floor. The class means can act the other way, stacking variance onto a few strong dimensions (right). The two effects separate exactly. Each rung's covariance is the pooled within-class covariance $W$ plus the covariance $B$ of the class means, and its log PR splits into a within-class term and a remainder $c$ that vanishes when the means coincide (Appendix~\ref{app:methods}),
\[
  \log \PR(W + B) = \log \PR(W) + c(W, B).
\]
Taken on every rung of both ladders (the floor's random subsets keep their labels for the split), the slopes split $\delta$ into a within-class part and a class-mean part.

In every recording the clouds add more than the means take away (Figure~\ref{fig:mechanism}B). The within-class part, $+0.29$ to $+0.64$, exceeds $\delta$, and the class-mean part, $-0.08$ to $-0.35$, pulls it down. With the tuning inside each $45^\circ$ bin removed, about half of the within-class part (41 to 57\%) is each neuron's variance changing with the class, nearly linearly in its mean (a Fano-like law). The rest is correlation structure that changes with direction, the population form of stimulus-dependent noise correlations \citep{kohn2005,ponce2013,nejatbakhsh2023,zylberberg2016,franke2016}.

The lift is more than a byproduct of single-neuron tuning \citep{elsayed2017}: in seven of eight recordings it needs correlated fluctuations that change with direction. Shuffling each neuron's trials within its class, which keeps the class means and each neuron's variance and removes the correlations, turns the shift negative in seven of eight recordings. Giving every class the same pooled fluctuations keeps the correlations and removes their dependence on the class; it takes the within-class part to near zero and the shift negative in all eight (both with in-bin tuning removed). Projecting out the class means removes what a linear decoder reads, and linear decoders given the same removal, estimated within their training folds, fall to chance. The shift, now almost entirely within-class, still clears its permutation test ($p \le 0.025$) in every half of every recording (Appendix~\ref{app:methods}). Direction is written in the fluctuations too.

\section{Two measured terms account for the shift}
\label{sec:account}

Two terms computed from the covariances, with nothing fitted, account for the two parts: the fluctuations' overlap for the within-class part, the means' placement against them for the class-mean part. Suppose every pair of classes shares the same fraction $\sigma$ of a class's overlap with itself (mean field; Figure~\ref{fig:multipole}C shows the measured overlaps). Then $K$ classes multiply the fluctuations' PR by $K/[1 + (K-1)\sigma]$: by $K$ when they share nothing, by one when they share everything. The ladder grows $K$-fold in size, so the within-class part is about $\ln[K/(1+(K-1)\sigma)]/\ln K$, the overlap law. The overlap identity, in its finite-sample form at the rungs' sample counts, keeps what mean field leaves out: overlaps that depend on separation, and the floor's first-rung sampling. It gives the part with nothing fitted (held-out mean error $-0.003$ in imaging; Appendix~\ref{app:methods}).

The overlaps carry the same symmetry as the means. With equal class weights and overlaps that depend only on separation, the identity is a structure factor. An order that adds opposite directions in pairs then sees only the even harmonics, as a crystal with a two-atom basis extinguishes some reflections. The overlaps' odd-to-even ratio names the order the fluctuations favour in all 43 mouse and macaque recordings (Appendix~\ref{app:order}).

The class-mean part's sign is set by where the means sit against the fluctuations. Means along the strongest fluctuations stack variance onto them and lower PR; means along weak ones fill those dimensions and raise it. A declared ladder that starts with one class holds no class-mean variance on its first rung. On all 1{,}536 full-population imaging ladders that start that way, the class-mean part takes the sign of the remainder at the full stimulus set (Appendix~\ref{app:methods}). Three measured numbers decide the full-set remainder's sign: the means' variance relative to the fluctuations', $\beta = \mathrm{Tr}\,B/\mathrm{Tr}\,W$; their alignment $a$ with the fluctuations; and their dimension ratio, $\PR(W)/\PR(B)$. In these numbers, the remainder is negative exactly when
\begin{equation}
  \frac{\PR(W)}{\PR(B)} > 1 + \frac{2\,(1 - a)}{\beta}, \qquad a = \frac{\mathrm{Tr}\,WB/\mathrm{Tr}\,B}{\mathrm{Tr}\,W^2/\mathrm{Tr}\,W},
  \label{eq:sign}
\end{equation}
(Appendix~\ref{app:methods}). When the fluctuations are at least as strong along the means as their variance-weighted average ($a \geq 1$), the means lower PR whenever $\PR(W) > \PR(B)$.

In imaging the means lie along the fluctuations ($a = 0.96$ to $2.1$), as gain fluctuations along the tuning would place them \citep{goris2014,lin2015,arandia2016}. The fluctuations span 73 to 173 dimensions against the means' 4 to 6, and the remainder is negative in all eight. With the overlap law, the class-mean term carried through the rungs reconstructs the shift's size across the eight recordings from three mice ($r = 0.88$; Figure~\ref{fig:twoterm}). Every imaging shift is positive, so an always-positive rule gets every sign right; only the size tests the account here. The sign is tested where it changes.

\section{Where the sign changes: a second laboratory and macaque V1}
\label{sec:allen}

Allen Neuropixels recordings \citep{allen2019,siegle2021}, spike counts from awake mice in populations of 51 to 421 units, keep the lift and reverse what the class means do. The shift is positive in all 32 VISp mice (mean $+0.20$; Appendix~\ref{app:allen}).

Each direction class pools five temporal frequencies, whose tuning would count as fluctuation. With that tuning removed, as for every split and term below (Appendix~\ref{app:allen}), the within-class part is positive, as in imaging (mean $+0.10$, positive in 30 of 32 mice), but the class-mean part is positive in 27 of 32. Eq.~\ref{eq:sign} says why. Allen's class means lie off the fluctuations' largest modes ($a = 0.24$ to $0.34$ against imaging's $1.08$ to $1.21$, medians at matched sizes, 64 to 256 units), which puts the threshold above $\PR(W)/\PR(B)$: the means fill weak modes and raise PR. Swapping only the median $a$ swaps the sign at every size.

The datasets differ in what their largest fluctuations are: in imaging they lie along the tuning, as gain does; in Allen they are shared, untuned and partly the animal's state. One of the three largest modes tracks running speed in 29 of 32 mice, and one tracks pupil area in 25 of the 26 with eye tracking ($|r| \geq 0.3$), as locomotion and arousal modulate visual cortex \citep{niell2010,vinck2015,akella2025}. Remove these three modes, estimated on the other half of the trials, and the class-mean part turns negative in 29 of 32 mice, against 3 on the same halves before, as in imaging (Appendix~\ref{app:allen}).

\begin{figure}[t]
\centering
\includegraphics[width=\linewidth]{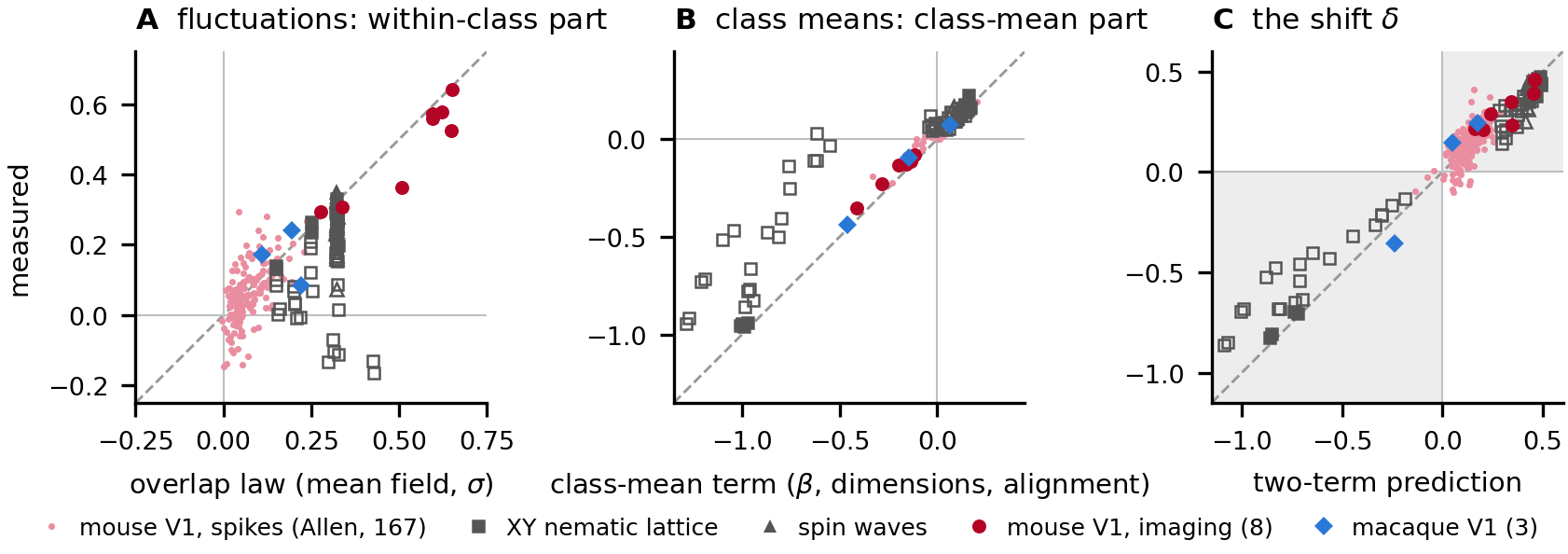}
\caption{The two terms against the measured parts, the within-class term in mean field. (\textbf{A})~Within-class part against the overlap law; (\textbf{B})~class-mean part against Eq.~\ref{eq:sign}'s term carried through the rungs; (\textbf{C})~the shift against their sum, signs agreeing in the shaded quadrants. Open markers: sampling term $s > 0.05$ (Appendix~\ref{app:methods}).}
\label{fig:twoterm}
\end{figure}

Across the 167 populations the shift itself changes sign, negative in 16 with the tuning removed. Computed on the same trials with the finite-sample overlap identity, the two terms give 14 of the 16 negative signs and 149 of the 151 positive ones (163 of 167; $r = 0.94$; Appendix~\ref{app:allen}). Mean field gets only 2 of the 16 negatives (Figure~\ref{fig:twoterm}; 152 of 167, against 151 for always positive). Where classes share most of their fluctuations, the within-class part's sign depends on which pairs overlap, the structure mean field averages into one $\sigma$.

Utah multielectrode arrays in anesthetized macaque V1 \citep{smith2008,kohn2016crcns} add a third laboratory and a second species. The shift is positive in two animals and negative in the third ($-0.35$), and the two terms give all three signs (Appendix~\ref{app:allen}). In the negative array the means carry twice the fluctuations' variance and lie along them, so the class-mean part ($-0.44$) outweighs a positive within-class part. Where class means sit against the fluctuations decides whether they raise the exponent or lower it.

\section{Systems with a known answer}
\label{sec:observable}

The recordings test the account where its sign changes; systems with a known answer test whether the shift reads it back. The simplest answer is zero. A rotation axis gives no shift beyond its spread over seeds in the invariant layers of a $C_4$-equivariant network, built to remove rotation, while a digit axis there does (Appendix~\ref{app:cnn}). A declared path can also withhold structure. In the critical 2D Ising model, adding configurations by increasing $|m|$ keeps the dominant mode, the magnetization $m$, out of the first rungs, and the shift is negative on every lattice tested, from $-0.50$ at $L = 32$ to $-0.32$ at $L = 256$ (Appendix~\ref{app:lattice}).

The XY model shows the two terms at work with the symmetry held fixed. Its spins are planar angles on a lattice, coupled by a nematic term that prefers aligned axes and a weak polar term that prefers aligned arrows, the analogues of orientation and direction. Configurations are sorted into eight classes by director angle, an orbit on $\mathbb{RP}^1$. Where polar order holds ($T \le 0.85$), each configuration's random polar sign is a few-dimensional fluctuation that turns with the director class, and both parts and the shift are positive. Above it the fluctuations span tens to hundreds of dimensions, the class means, a single harmonic, concentrate the variance, and the mean shift is negative from $T = 0.9$ to $1.1$ (Appendix~\ref{app:lattice}). The two terms give the sign in all 72 lattice and spin-wave ensembles (Figure~\ref{fig:twoterm}). The classes lie on the same orbit at every temperature; the fluctuations decide the sign.

GPT-2 completes prompts such as `Then, Alice and Mary went to the office. Alice gave a computer to' with the indirect object, Mary. Path patching found the eleven name-mover and backup heads of the circuit that does this \citep{wang2023ioi}. With prompts as samples, each attention head's output as features and the indirect object's name as the axis (eight names), the shift's permutation $z$ ranks those heads at AUC $0.93$ among all 144 (ten seeds), with no classifier trained: a circuit head outranks a randomly chosen non-circuit head 93 times in 100. A trained linear classifier ranks them at $0.95$, and the heads' class-mean variance share, also classifier-free, at $0.96$ (Appendix~\ref{app:llm}). The ranking reads how the name axis organizes each head's write; which heads the output uses takes an intervention.

\section{Discussion}
\label{sec:discussion}

These results change how a dimensionality exponent should be reported: with its path, its floor and its shift. Random sampling is itself a probe, and it returns the floor; structure beyond the floor appears only along a declared axis, which sets its size and sector. There is no probe-free dimensionality exponent. For neuroscience, the random-subsample exponents of these 18 recordings differ mostly in where their rungs sit against their effective dimensions, and other covariance statistics carry floors of their own (Appendix~\ref{app:allen}). The shift's split weighs stimulus-dependent fluctuations against the class means, whose effect reversed in a second laboratory where the largest fluctuations track the animal's state.

\paragraph{For machine learning.} Dimensionality profiles across a language model's layers are read as stages of processing \citep{valeriani2023,cheng2025}. With prompts as samples and a layer's units as features, a topic axis, averaged over class orders, lifts every layer of four Pythia models: each topic adds directions the others occupy weakly, and more so in late layers (Appendix~\ref{app:llm}). In all four the late lift survives matching final tokens per class, which move estimated dimensionality \citep{osman2026}. Probes read where the class means sit \citep{alain2017,hewitt2019control}. With the topic means projected out, a probe falls to chance while the lift stays above its null at 28 of 33 layers of Pythia-2.8B: the topics differ in how their prompts vary as well as in where they sit, as directions do in cortex. GPT-2's head ranking (Section~\ref{sec:observable}) is this reading for one head at a time.

\paragraph{Scope and predictions.} The account is one of second moments. A zero shift says only that the declared ladder climbs like its floor: higher moments and variables off the path go unread, as scattering sees order only at the wavevectors it probes. Shift sizes compare within one design (Appendix~\ref{app:orient}). Two predictions follow for new data. In superficial macaque V1, with orientation maps and no direction map \citep{lu2010}, columns pool opposite directions, so mesoscale signals should lose what orientation-sorted blocks of neurons lose here (Appendix~\ref{app:rg}). The overlaps' odd-to-even ratio, measured on some trials of a new preparation, should name the order the fluctuations favour on the rest: adjacent directions above one, opposite pairs below (Appendix~\ref{app:order}).

\newpage
\acks{We thank the three reviewers and the area chair, whose comments sharpened the paper. We thank Francisco Acosta
and the NeurReps organizers for their help with the proceedings. We are grateful to Carsen Stringer, Marius Pachitariu,
Adam Kohn, Matthew Smith, their colleagues and the Allen Institute for making their recordings public.}

\bibliography{theta}
\clearpage

\appendix
\raggedbottom

\begin{table}[H]
\centering
\small
\begin{tabular}{@{}l>{\raggedright\arraybackslash}p{0.66\linewidth}l@{}}
\toprule
 & What the appendix shows & Main text \\
\midrule
\ref{app:identity} & The split $\tobs = \tfloor + \delta$ is exact for ladders that share rung sizes & Eq.~\ref{eq:decomp} \\
\ref{app:random} & Random-stimulus ladders sit at their floor; the floor and the shift's threshold are sampling crossovers & Sections~\ref{sec:decomp},~\ref{sec:results} \\
\ref{app:orient} & Per-recording shifts with resampling intervals, permutations, origins and random class orders & Section~\ref{sec:results} \\
\ref{app:neuron} & Neighbouring-neuron ladders under both centerings & Section~\ref{sec:results} \\
\ref{app:order} & Accumulation order: the fluctuation overlaps' harmonics name the order the fluctuations favour & Section~\ref{sec:account} \\
\ref{app:lattice} & Ising and XY lattices: samplers, shifts and the two-term account in 72 ensembles & Section~\ref{sec:observable} \\
\ref{app:rg} & Coarse-graining: which blocks of neurons keep which sector, and the mesoscale prediction & Section~\ref{sec:discussion} \\
\ref{app:allen} & Allen and macaque populations: tuning, temporal frequency, state modes, the two terms, axis screening & Sections~\ref{sec:allen},~\ref{sec:discussion} \\
\ref{app:llm} & Language models: order averages, depth, final tokens, benchmarks, probes, the GPT-2 circuit & Sections~\ref{sec:observable},~\ref{sec:discussion} \\
\ref{app:cnn} & An equivariant network whose invariant layers read no rotation & Sections~\ref{sec:decomp},~\ref{sec:observable} \\
\ref{app:methods} & Datasets, estimator, inference, harmonics, the covariance split and its controls, decoders, the overlap identity, Eq.~\ref{eq:sign} & Sections~\ref{sec:results}--\ref{sec:account} \\
\ref{app:wishart} & Wishart derivation of Eq.~\ref{eq:null} & Section~\ref{sec:decomp} \\
\ref{app:worked} & One recording through \texttt{rung}, and what each option returns & Section~\ref{sec:intro} \\
\bottomrule
\end{tabular}
\caption{What each appendix shows, and the part of the main text it supports.}
\label{tab:roadmap}
\end{table}

\section{The identity}
\label{app:identity}
Write $y_k = \log \PR^{(k)}$ for the observed ladder,
$f_k = \log \PR^{(k)}_{\mathrm{null}}$ for the matched-null ladder
(the mean log-PR across null draws; its slope is the floor), and $x_k = \log N_k$.
Ordinary least squares is linear in the response, so
$\hat\beta_{y|x} = \hat\beta_{f|x} + \hat\beta_{(y-f)|x}$, i.e.
$\tobs = \tfloor + \delta$. The decomposition holds for any ladder and
any null, and for any fixed regression weights shared by both arms; the physical content lies in the choice of null and axis.

\section{Random-stimulus zoom, and the two sampling crossovers}
\label{app:random}
Across all 18 recordings and stimulus conditions the random ladder sits at its floor (Table~\ref{tab:random}).
\begin{table}[h]
\centering
\small
\begin{tabular}{lcccc}
\toprule
Type & $n$ rec. & Mean $\tobs$ & Mean floor \% & Mean $\delta_{\mathrm{rand}}$ \\
\midrule
Drifting gratings & 3 & 0.203 & 97.4\% & $+0.007$ \\
Local gratings & 3 & 0.284 & 102.1\% & $-0.005$ \\
Low-contrast gratings & 2 & 0.340 & 101.7\% & $-0.006$ \\
Static, orientation-biased & 2 & 0.135 & 98.4\% & $+0.001$ \\
Spontaneous & 8 & 0.555 & 100.7\% & $-0.003$ \\
\midrule
\textbf{All} & \textbf{18} & 0.381 & \textbf{100.3\%} & $\mathbf{-0.002}$ \\
\bottomrule
\end{tabular}
\caption{Random-stimulus subsampling. The matched floor tracks the
observed exponent in every recording (mean shift $-0.0017$, largest
single-recording magnitude $0.020$). Spontaneous sessions have larger
$\tobs$ because their effective rank differs, with identical floor
domination.}
\label{tab:random}
\end{table}
The floor is a crossover. Rung by rung, the local slopes of the random ladders in the 18 recordings fall with $n/d_{\mathrm{eff}}$ ($d_{\mathrm{eff}}$ the PR at the full stimulus set) in the shape of the iid form $1/(1 + n/d_{\mathrm{eff}})$ of Eq.~\ref{eq:null}, about a quarter above it (correlation $0.99$ over 108 rung pairs, pooled; median ratio to the iid form $1.26$; all 18 ladders fall at every rung). So a ladder inherits its floor from where its rungs sit: restricted to rungs of at least 500 stimuli, the random ladder's slope ($0.052$--$0.138$) matches the direction ladder's floor ($0.055$--$0.138$) in all eight grating recordings.

The shift has a threshold of its own. The GT3 direction ladder on random
subsets of the stimuli (rungs below ten samples skipped; ten subsets
per size up to 500, three to six above) gives $\delta = +0.03 \pm 0.11$ (mean $\pm$ SD over subsets),
$+0.02 \pm 0.12$ and $-0.00 \pm 0.06$ at 50, 100 and 200 stimuli, then
$+0.08 \pm 0.06$ at 300, $+0.15 \pm 0.03$ at 1{,}000 and $+0.234$ at the
full 4{,}442.

The covariance split of Section~\ref{sec:sign} says why.
Each direction class has an effective fluctuation dimension
$d_w = 18$ to $31$ (its within-class PR over all its presentations, 18
to 29, inverted through Eq.~\ref{eq:null}). With fewer presentations
per class than that, a class's fluctuations are sampling-limited: the
within-class part is set by the degrees of freedom each subset loses to
its class means (Appendix~\ref{app:methods}), negative ($-0.25$ at 6
presentations per class, $-0.08$ at 25) and cancelled by the class-mean
part. It crosses zero between 25 and 38 presentations per class, near
$d_w$, and then grows ($+0.10$, $+0.22$, $+0.30$, $+0.36$ at 62, 125,
250, 555 per class) while the PR of a single class climbs toward $d_w$
along the crossover of Eq.~\ref{eq:null} ($15$, $19$, $22$, $25$ against
the crossover's $18$, $21$, $23$, $25$).

The
shift reads how classes differ in their fluctuations, which on GT3
becomes visible as each class is sampled beyond its own fluctuation
dimension: there the crossover of Eq.~\ref{eq:null}, applied to one class
in the declared ladder and to the pooled fluctuations in its floor, sets the
threshold and the climb after it, as the same crossover applied to the
population sets the floor. The same threshold holds in every grating recording: the within-class part turns positive for good
once each class holds 20 to 45 presentations, 0.46 to 1.2 times that recording's own fluctuation dimension (within a factor of
two in six of eight; the onset rises with the dimension, Spearman $0.79$), and the low-contrast recordings, whose fluctuations
span the most dimensions ($78$ and $97$), need the most ($42$ and $45$). The finite-sample form of the overlap identity at each subset's sample counts (Appendix~\ref{app:methods}) tracks the eight recordings' curves ($r = 0.99$ over 64 cells) and puts their zero crossings within 25\% in six of eight.

\section{Direction-aligned zoom: per-recording results, bootstrap, and shuffle control}
\label{app:orient}
\begin{figure}[h]
\centering
\includegraphics[width=0.78\linewidth]{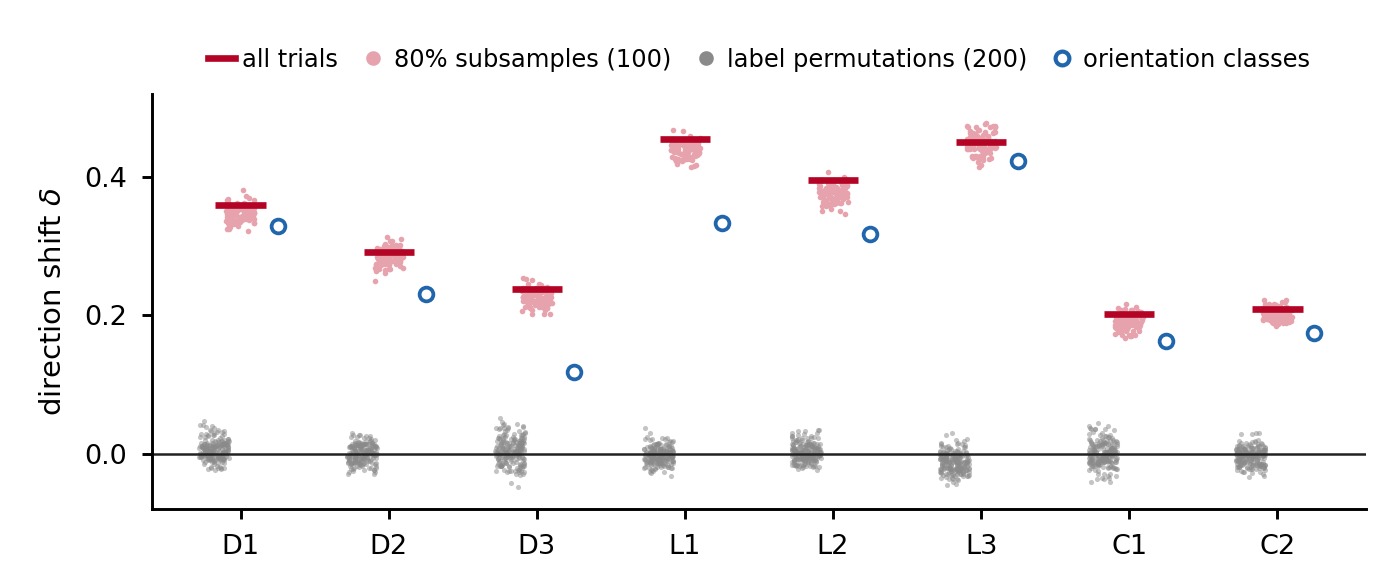}
\caption{Direction-aligned shift per grating recording. Bar: all trials (Table~\ref{tab:orient}); pink: 100 estimates from 80\% of each class's trials drawn without replacement, floor redrawn in each; gray: 200 label permutations at the same rung sizes; open circles: orientation classes. Every shift exceeds all its permutations, and the subsamples' median sits below it in all eight.}
\label{fig:delta_recordings}
\end{figure}

Two resampling tests separate uncertainty from alignment (Figure~\ref{fig:delta_recordings}). The
resampling interval draws 80\% of each class's trials without
replacement (100 replicates, percentile 95\% interval, floor redrawn in
every replicate); label permutation at fixed rung sizes (floor
unchanged) tests whether the shift depends on the class assignment.
The grating sessions span three animals, all with positive mean
shifts; animal-level inference rests on the 32 single-animal sessions
of Section~\ref{sec:allen}.
\begin{table}[h]
\centering
\small
\begin{tabular}{lcccccc}
\toprule
Recording & Neurons & $\delta_{\mathrm{dir}}$ & 95\% interval & $\bar\delta_{\mathrm{shuffled}}^{\mathrm{dir}}$ & $\delta_{\mathrm{ori}}$ & $\bar\delta_{\mathrm{shuffled}}^{\mathrm{ori}}$ \\
\midrule
Drifting GT1 & 20{,}616 & $+0.360$ & $[0.326, 0.368]$ & $+0.006$ & $+0.328$ & $-0.012$ \\
Drifting GT2 & 14{,}034 & $+0.292$ & $[0.265, 0.307]$ & $+0.001$ & $+0.231$ & $-0.002$ \\
Drifting GT3 & 11{,}311 & $+0.237$ & $[0.204, 0.248]$ & $-0.009$ & $+0.118$ & $+0.000$ \\
Local GT1 & 22{,}482 & $+0.455$ & $[0.417, 0.457]$ & $-0.003$ & $+0.332$ & $+0.010$ \\
Local GT2 & 17{,}750 & $+0.395$ & $[0.352, 0.396]$ & $-0.006$ & $+0.318$ & $-0.013$ \\
Local GT3 & 15{,}640 & $+0.450$ & $[0.423, 0.474]$ & $-0.021$ & $+0.422$ & $-0.009$ \\
Low-contrast GT1 & 22{,}565 & $+0.202$ & $[0.170, 0.210]$ & $+0.001$ & $+0.163$ & $-0.001$ \\
Low-contrast GT2 & 16{,}159 & $+0.209$ & $[0.188, 0.218]$ & $+0.005$ & $+0.174$ & $+0.010$ \\
\midrule
Static TX40$^{\dagger}$ & 17{,}631 & $+0.003$ & --- & $-0.008$ & $+0.008$ & $-0.002$ \\
Static TX42$^{\dagger}$ & 18{,}982 & $+0.064$ & --- & $-0.024$ & $+0.056$ & $-0.032$ \\
\bottomrule
\end{tabular}
\caption{Direction-aligned shifts (point estimate and resampling
interval), shuffled-label means under both binnings (five permutations
per cell), and orientation-binned variants. Every point lies inside its
interval, and the subsample medians sit slightly below the points (Figure~\ref{fig:delta_recordings}). Figure~\ref{fig:ladder}
and the three-decimal split of Section~\ref{sec:results} use an
independent set of ten floor draws (GT3: $+0.240$). Resampling with
replacement instead inflates the centers (GT2 low contrast: $0.273$
$[0.237, 0.318]$ against the point $0.209$), because duplicated trials
lower early-rung PR and steepen the ladder, so the intervals draw without replacement.
Against 200 label permutations every grating cell under both binnings
exceeds all permuted values ($p = 1/201$; null mean within $\pm 0.013$, SD $0.011$--$0.018$).
$^{\dagger}$Orientation-biased static ensembles: the stimulus variable
spans $\sim 4^\circ$, so their eight classes are nearly one stimulus.}
\label{tab:orient}
\end{table}

Rotating the binning origin through all eight cyclic starts and
reversing the traversal direction preserves a positive shift in all
128 recording-variant cells; the magnitude varies with origin
(per-recording SD $0.04$--$0.08$). Random orders of the eight whole classes (40 per recording, each with its own floor) return shifts of the same size, positive in all 320 (recording means $+0.13$ to $+0.41$), and the angular order sits at their 15th to 90th percentile, unrelated to the quadrupole-to-dipole ratio ($\rho = -0.07$): the direction partition carries the shift, and the path through it moves the shift by about one random-order SD ($0.04$--$0.09$). The baseline angular ladders start at zero degrees; order effects with a physical reading, and their dependence on the origin, are the targeted contrasts of Appendix~\ref{app:order}.

Ladder-design variants preserve
the conclusion in every grating recording: dropping the first rung,
dropping the last, and a coarse schedule ($1, 2, 4, 8$ classes) all
leave the shift positive (24 of 24 recording-variant
cells), and dropping the first rung raises it. The slope-free deficit area
$\int [\log \PR_{\mathrm{floor}} - \log \PR_{\mathrm{aligned}}]\,
d\log n$ is positive in all eight recordings: aligned rungs sit
below the floor throughout the nested ladder, the same structure
summarized without a fitted slope.

\section{Spatial neuron ladder}
\label{app:neuron}
The observed ladder takes, at each rung size $n$, the $n$ nearest
neighbours of a random anchor cell (recorded positions), with PR averaged over
50 anchors before the logarithm; the floor uses random same-size subsets with the identical
estimator. The control column repeats the analysis with random subsets
in both arms, which is exchangeable and must vanish in expectation. Which neurons a ladder samples, including how it covers clusters, also changes factor-analysis shared dimensionality \citep{williamson2016}; this ladder measures centered PR against same-size random subsets.
\begin{table}[h]
\centering
\small
\begin{tabular}{lcccc}
\toprule
Recording & $\theta_{\mathrm{spatial}}$ & $\theta_{\mathrm{floor}}$ & $\delta_{\mathrm{spatial}}$ & Control \\
\midrule
Drifting GT1 & $0.318$ & $0.343$ & $-0.026$ & $+0.006$ \\
Drifting GT2 & $0.278$ & $0.335$ & $-0.057$ & $+0.018$ \\
Drifting GT3 & $0.346$ & $0.342$ & $+0.004$ & $-0.004$ \\
Local GT1 & $0.447$ & $0.472$ & $-0.025$ & $+0.017$ \\
Local GT2 & $0.401$ & $0.472$ & $-0.070$ & $+0.005$ \\
Local GT3 & $0.379$ & $0.399$ & $-0.021$ & $+0.011$ \\
Low-contrast GT1 & $0.413$ & $0.418$ & $-0.005$ & $-0.003$ \\
Low-contrast GT2 & $0.403$ & $0.465$ & $-0.061$ & $+0.003$ \\
Static TX40 & $0.255$ & $0.325$ & $-0.070$ & $-0.012$ \\
Static TX42 & $0.271$ & $0.267$ & $+0.004$ & $-0.011$ \\
\midrule
Mean (SE) & & & $-0.033$ $(0.009)$ & $+0.003$ \\
\bottomrule
\end{tabular}
\caption{\label{tab:spatial}Spatially contiguous neuron subsets carry slightly less
dimensionality than random subsets (8 of 10 negative), while the
exchangeable control has mean near zero ($+0.003$).}
\end{table}

The neuron ladders center each presentation across the sampled units, the stimulus ladders each unit across the sampled presentations. Centering each unit instead, on the same subsets (ten draws per size; Table~\ref{tab:spatial} uses 50), raises GT3's spatial exponent from $0.354$ to $0.411$ while its shift stays near zero ($-0.005$, $-0.001$); across the ten recordings the mean shift is $-0.035$ and $-0.045$, negative in 8 of 10 under either centering. The centering moves the exponent, and the floor moves with it.

\section{Accumulation-order experiments}
\label{app:order}
Sixteen direction classes ($22.5^\circ$) accumulated two per rung at
closely matched rung sizes (each arm with its own floor), in two
orders: sequential (adjacent classes) and
antipodal-paired (each rung adds a class and its $180^\circ$ partner,
completing one orientation class per rung). From the declared origin the sequential order climbs faster in every recording (Table~\ref{tab:order}). The split of Table~\ref{tab:split} gives the effect, paired minus sequential, two parts: the within-class part favours the sequential order in all eight, the class-mean part the paired order in five ($+0.06$ to $+0.14$), is near zero in the two low-contrast recordings and favours the sequential order in Drifting GT3.

\paragraph{Across origins.} Starting the two orders from each of the sixteen classes moves the total and keeps the large parts. Averaged over origins the total is $-0.046$ to $+0.041$ (negative in six recordings; negative at 6 to 12 of the 16 origins), while in the three localized recordings the within-class part favours the sequential order ($-0.15$ to $-0.20$) and the class-mean part the paired order ($+0.11$ to $+0.17$) from all sixteen.

Averaged over origins, the within-class part is an exact function of the sixteen classes' overlap matrix (Appendix~\ref{app:methods}); computed from one temporal half of the trials, it predicts the other half's order effect to a median $0.005$ ($r = 0.996$). Across these eight recordings the effect falls as the adjacent-minus-antipodal overlap grows ($r = -0.997$), a summary of that matrix: overlap matrices whose summaries agree within $0.02$ can still differ in the order they favour, by up to $0.066$ and in sign.

\paragraph{One ratio across preparations.} As a function of separation the fluctuation overlaps split into even and odd harmonics. With equal class weights and a separation-only kernel, an order that adds opposite directions in pairs reads only the even ones, the extinction rule of a two-atom crystal basis (Appendix~\ref{app:methods}).

With eight classes (twelve in macaque), take the ratio of odd to even harmonic power (summed squared coefficients) of the fluctuation overlaps, the split-half shape overlap of two classes averaged by separation. It separates the order the fluctuations favour (the within-class part of paired minus sequential) in all 43 recordings, mice and arrays: above one in the three localized recordings ($1.9$ to $3.2$), where the adjacent order climbs faster, and below one where the paired order wins, in the drifting and low-contrast recordings ($0.24$ to $0.90$), the 32 Allen VISp populations (median $0.012$) and the three macaque arrays ($0.013$ to $0.020$). Within imaging the ratio follows the direction-selective fraction (Spearman $0.95$).

The ratio and the adjacent-minus-antipodal overlap summarize one matrix, and the ratio classifies better (the overlap alone, 41 of 43); in a new preparation it gives a first reading of the order its fluctuations favour, and the weighted overlap identity decides (Appendix~\ref{app:methods}). In the equivariant network of Appendix~\ref{app:cnn}, whose opposite rotations share more of their fluctuations than neighbours, it names the paired order, which climbs faster.

In Allen opposite directions share more fluctuation than neighbours in all 32 VISp populations (in five of the eight imaging recordings neighbours share more), and the fluctuations favour the paired order with the sign the overlaps predict in all 32 ($r = -0.84$); in all three macaque arrays opposite directions share more fluctuation than neighbours and the fluctuations favour the paired order ($+0.045$ to $+0.086$), as the overlaps predict. At eight classes the full-field recordings favour the paired order slightly ($+0.004$ to $+0.13$), while at sixteen classes from the declared origin all eight favour the sequential order: the class count is part of the design.
\begin{table}[h]
\centering
\small
\begin{tabular}{lccc}
\toprule
Recording & $\delta_{\mathrm{sequential}}$ & $\delta_{\mathrm{paired}}$ & Order effect \\
\midrule
Drifting GT1 & $+0.379$ & $+0.339$ & $-0.039$ \\
Drifting GT2 & $+0.308$ & $+0.237$ & $-0.072$ \\
Drifting GT3 & $+0.254$ & $+0.125$ & $-0.129$ \\
Local GT1 & $+0.469$ & $+0.341$ & $-0.128$ \\
Local GT2 & $+0.409$ & $+0.340$ & $-0.069$ \\
Local GT3 & $+0.463$ & $+0.435$ & $-0.028$ \\
Low-contrast GT1 & $+0.226$ & $+0.163$ & $-0.063$ \\
Low-contrast GT2 & $+0.218$ & $+0.184$ & $-0.034$ \\
\bottomrule
\end{tabular}
\caption{Accumulation order at matched rung sizes, from the declared origin. The paired
(quotient-respecting) order climbs less steeply than the sequential
order in 8 of 8 recordings; the within-class fluctuations carry the preference
(Table~\ref{tab:split}); other origins in the text.}
\label{tab:order}
\end{table}

\section{Ground-truth lattice models}
\label{app:lattice}
\begin{figure}[h]
\centering
\includegraphics[width=\linewidth]{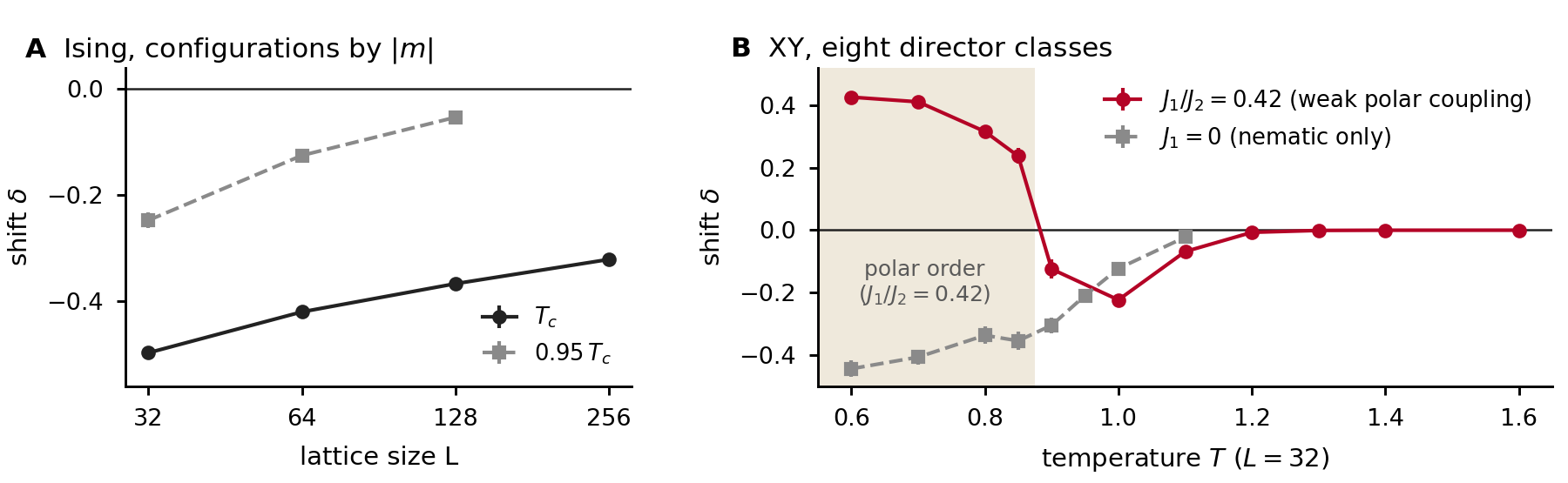}
\caption{Lattices of known structure. (\textbf{A})~Ising: adding configurations by increasing $|m|$ gives a negative shift at $T_c$ on every lattice size (chain means $\pm$ SE), collapsing off criticality. (\textbf{B})~XY with eight director classes: with a weak polar coupling (red) the shift is positive while polar order holds, negative as it melts and near zero above; with no polar coupling (gray) it is negative at every temperature tested.}
\label{fig:lattices}
\end{figure}

\paragraph{2D Ising (scalar order parameter).}
Wolff-cluster sampling, 200 configurations per chain, ten-draw floors. Ladders are nested: the rungs hold the first $10$, $25$, $50$, $75$ and $100\%$ of the configurations in order of increasing magnetization ($m$ or $|m|$); floors use random same-size configuration subsets. At the critical temperature $T_c$ each size has 16 independent chains (8 at $L = 256$), and Binder cumulants certify the sampler: $U(T_c) = 0.609$, $0.615$, $0.608$, $0.614$ at $L = 32$, $64$, $128$, $256$, against the universal $0.611$.

With the centered estimator used on the neural data, sorting by $|m|$ (Figure~\ref{fig:lattices}A) gives $\delta(T_c) = -0.497 \pm 0.008$ (chain means $\pm$ SE), $-0.420 \pm 0.009$, $-0.367 \pm 0.008$ and $-0.321 \pm 0.007$ at $L = 32$, $64$, $128$ and $256$, negative in every chain: the first rungs hold only configurations near $m = 0$, which lack the dominant fluctuation mode, the magnetization itself, so they start at high PR.

The magnitude falls as $L^{-0.21}$ (chain-bootstrap 95\% interval $[0.19, 0.23]$), following the magnetization mode's share of the configuration variance, $\langle m^2\rangle \sim L^{-0.245}$, monotonically but slightly more slowly ($|\delta| \approx 1.1\,\langle m^2\rangle$ at every size); the smaller sizes predict $-0.31$ at $L = 256$, where $-0.321$ is measured. A spectral account with nothing fitted, the magnetization mode plus the bulk fluctuations measured on each rung, reproduces the shift within $0.03$ at every size ($-0.51$, $-0.45$, $-0.39$, $-0.33$). Off criticality the shift collapses ($-0.247$, $-0.125$, $-0.054$ at $0.95\,T_c$, $L = 32$--$128$).

The ladder conditions on sampled configurations and couples no field to the system, so $\delta$ is a conditioning observable with a scaling of its own: at $T_c$ it follows the share of the variance that the conditioning removes.

\paragraph{2D XY nematic (orbit-valued order parameter).}
Planar spins $\phi_i$ on an $L \times L$ lattice have the energy
\[
  E = -\sum_{\langle ij\rangle}\big[J_2\cos 2(\phi_i - \phi_j) + J_1\cos(\phi_i - \phi_j)\big],
\]
whose nematic term ($J_2 = 1$) prefers aligned axes and whose polar term prefers aligned arrows; $J_1/J_2 = 0.42$ (a weak polar coupling) or $J_1 = 0$. Sampling is a reflection-Wolff cluster hybrid; hot and ordered starts reach the same energy at every temperature quoted, and each ensemble holds 40 independent chains of five samples, so every director bin fills and no chain is selected. Configurations are binned by the global director angle (the phase of the quadrupole order parameter, on $\mathbb{RP}^1$) into eight classes and accumulated as in the cortical ladder; the estimator is centered.

At $L = 32$ (Figure~\ref{fig:lattices}B) the shift is positive on the polar-ordered side, $+0.43$, $+0.41$, $+0.32$ and $+0.24$ at $T = 0.6$, $0.7$, $0.8$ and $0.85$ (polar order $0.91$ at $0.7$, $0.84$ at $0.8$), and negative from $T = 0.9$, as polar order melts: $-0.12$, $-0.22$ and $-0.07$ at $T = 0.9$, $1.0$ and $1.1$ (the peak of the nematic susceptibility), near zero above; at $L = 64$, $-0.18$ at $T = 1.0$ and $-0.014$ at $1.1$; for $J_1 = 0$ negative at every temperature tested ($-0.44$ at $T = 0.6$ to $-0.02$ at $1.1$). Permuted labels average within $0.011$ of zero.

On the positive side each configuration's random polar sign, $\pm$ the polar order, is a large within-class fluctuation in a few dimensions whose direction turns with the director class: the fluctuations span few dimensions (PR $5.5$ to $5.7$ at $T = 0.7$), so the class means spread them, and both parts are positive (temperature means: within-class $+0.17$ to $+0.27$, class-mean $+0.07$ to $+0.17$). With the nematic features alone the same ensembles read $-0.84$ ($T = 0.7$).

On the negative side the class means span about two dimensions, one harmonic, against fluctuations of PR 150 to 500, and the shape overlap of two classes' fluctuations falls with separation to a plateau near $0.5$: about half of each class's fluctuation does not depend on the class. The within-class part is sampling-limited until a class holds many times its fluctuation dimension: at $L = 16$, where that dimension converges (PR $108$), it rises from $-0.09$ to $+0.25$ as configurations per class grow from 25 to 1600, reaching the overlap law ($\sigma = 0.54$), while the class-mean part deepens to $-0.95$ and the shift settles at $-0.70$. A Gaussian spin-wave model with the same order and director classes, whose fluctuations rotate with the director (overlap $1$, $0.85$, $0.51$, $0.17$, $0.03$ by separation), reads $+0.35$ and $+0.43$ at 25 and 200 configurations per class, positive in eight of eight seeds.

The two-term account of Section~\ref{sec:account}, computed from each ensemble with nothing fitted, gives the sign of the shift in all 72 ensembles: 24 on the negative side ($L = 16$ and $32$, 25 to 1600 configurations per class), 32 on the polar-ordered side ($T = 0.6$ to $0.85$; predicted $+0.48$ to $+0.31$ against measured $+0.43$ to $+0.24$) and the 16 spin-wave ensembles. Its size converges where a class holds many times its fluctuation dimension ($L = 16$, 1600 per class: $-0.74$ and $-0.72$ predicted against $-0.70$ and $-0.70$).

With fewer samples the within-class part sits below the overlap law, by $0.06$ to $0.16$ on the polar side at 25 configurations per class (temperature means). The floor's first rungs are under-sampled there, and the finite-sample form of the overlap identity (Appendix~\ref{app:methods}) gives that part in all 72 ensembles ($r = 0.99$, mean error $0.001$). The class-mean part moves the other way, and the sign holds at every size while the shift's size converges with the samples ($L = 16$, $T = 1.0$, two-ensemble means: $-0.27$ at 25 per class, $-0.70$ at 1600). The classes lie on the same orbit at every temperature, and the fluctuations around them decide the sign.

\section{Coarse-graining on real data}
\label{app:rg}
Coarse-graining a neural population into blocks of neurons goes back to \citet{meshulam2019}, who merge maximally correlated pairs; here $K$ neurons are averaged into single units, the blocks sorted by preferred direction (the phase of the first angular harmonic) or assigned at random, and the direction ladder is rerun at each $K$. On GT3, direction-sorted blocks keep the shift at every block size ($+0.20$ to $+0.27$ for $K = 2$ to $32$, against $+0.235$ ungrained) while random blocks mix tuning and lose it ($+0.027$ at $K = 32$, one floor seed). The sorting preserves it: with preferences, rates and reliabilities estimated on one temporal half and the shift measured on the other, direction-sorted blocks keep more of the $K = 32$ shift than random blocks matched for rate and reliability in all eight recordings (GT3: $+0.22$ and $+0.30$ against $-0.01$ and $+0.05$, both ways round), and matched blocks keep less than plain random ones in all eight.

Which neurons a block pools decides which sector survives. All eight grating
recordings are grained with four block types (Table~\ref{tab:graining}; $K = 1, 4,
16, 32$, two floor seeds averaged): blocks sorted by preferred
\emph{orientation} (the phase of the second harmonic), which pool
opposite directions; blocks sorted by preferred direction; random
blocks; and anatomical blocks, $k$-means clusters of the neurons'
imaging-plane positions. The shift is positive in all 128
recording-block-size cells.

Orientation-sorted blocks multiply the
quadrupole-to-dipole balance $b/c$ of the class means by $5.8$ to
$8.3$ at $K = 32$. They keep $0.77$ to $1.08$ of the ungrained shift in GT1, GT3 and the two low-contrast recordings, against $0.27$ to $0.52$ in the three dipole-dominant localized recordings and in GT2. GT2's class means are even-led ($b/c = 1.6$), but its shift rests on tuning inside each $45^\circ$ bin ($+0.292$, $+0.032$ with that tuning removed; Appendix~\ref{app:methods}): orientation-sorted blocks pool opposite directions, and at $K = 32$ they lower GT2's shift by $0.20$, and by $0.08$ once that tuning is removed first (one floor draw), so the tuning accounts for $0.12$ of the loss. Whether $b/c$ is above or below one sorts the other seven.

Direction-sorted blocks lower the balance to $0.50$ to $0.69$ of its
value and keep the most shift of the four in all eight. Anatomical
blocks behave as random ones: from $K = 16$ on their shift is within
$0.03$ of the random-block value in every recording and their balance
within $20\%$ of the ungrained one (at $K = 4$ they scatter on both
sides of it, by up to $0.23$). At these block sizes mouse anatomy
pools neurons as a random draw does. Orientation columns pool neurons by preferred orientation, as orientation-sorted blocks do, so mesoscale signals in cortex
without direction domains should lose the part of the shift that
orientation-sorted blocks lose here (Section~\ref{sec:discussion}).
\begin{table}[h]
\centering
\small
\setlength{\tabcolsep}{4pt}
\begin{tabular}{lcccccccc}
\toprule
 & & \multicolumn{4}{c}{shift kept at $K = 32$} & \multicolumn{3}{c}{balance $b/c$ at $K = 32$, relative} \\
\cmidrule(lr){3-6}\cmidrule(lr){7-9}
Recording & $\delta$, $K = 1$ & orient. & direct. & random & anat. & orient. & direct. & anat. \\
\midrule
Drifting GT1 & $+0.356$ & $0.86$ & $1.20$ & $0.90$ & $0.85$ & $6.18$ & $0.69$ & $1.01$ \\
Drifting GT2 & $+0.290$ & $0.31$ & $0.78$ & $0.55$ & $0.53$ & $8.33$ & $0.65$ & $1.15$ \\
Drifting GT3 & $+0.232$ & $0.96$ & $1.13$ & $0.06$ & $0.17$ & $7.27$ & $0.50$ & $1.05$ \\
Local GT1 & $+0.458$ & $0.34$ & $0.41$ & $0.28$ & $0.31$ & $7.44$ & $0.53$ & $1.19$ \\
Local GT2 & $+0.395$ & $0.27$ & $0.41$ & $0.03$ & $0.08$ & $7.63$ & $0.58$ & $1.12$ \\
Local GT3 & $+0.456$ & $0.52$ & $0.56$ & $0.35$ & $0.37$ & $6.79$ & $0.64$ & $1.20$ \\
Low-contrast GT1 & $+0.202$ & $1.08$ & $1.44$ & $0.64$ & $0.74$ & $5.84$ & $0.51$ & $0.94$ \\
Low-contrast GT2 & $+0.213$ & $0.77$ & $1.08$ & $0.54$ & $0.49$ & $6.61$ & $0.54$ & $1.02$ \\
\bottomrule
\end{tabular}
\caption{Graining by four block types (fraction of the ungrained shift kept at $K = 32$; class-mean balance $b/c$ at
$K = 32$ relative to $K = 1$; random blocks keep the balance within $0.95$ to $1.16$). Orientation-sorted blocks pool
opposite directions and amplify the quadrupole; direction-sorted blocks keep the dipole; anatomical blocks in mouse
track random ones.}
\label{tab:graining}
\end{table}

The shift is a resolution-dependent
observable \citep{recanatesi2022}, as the description a neuron's
computation receives depends on the resolution it is observed at
\citep{dehghani2026}, here along two zoom axes: graining, shown here, and sampling, which
has a threshold of its own (Appendix~\ref{app:random}).

\section{Allen Neuropixels and macaque arrays: per-population results}
\label{app:allen}
\begin{figure}[h]
\centering
\includegraphics[width=0.62\linewidth]{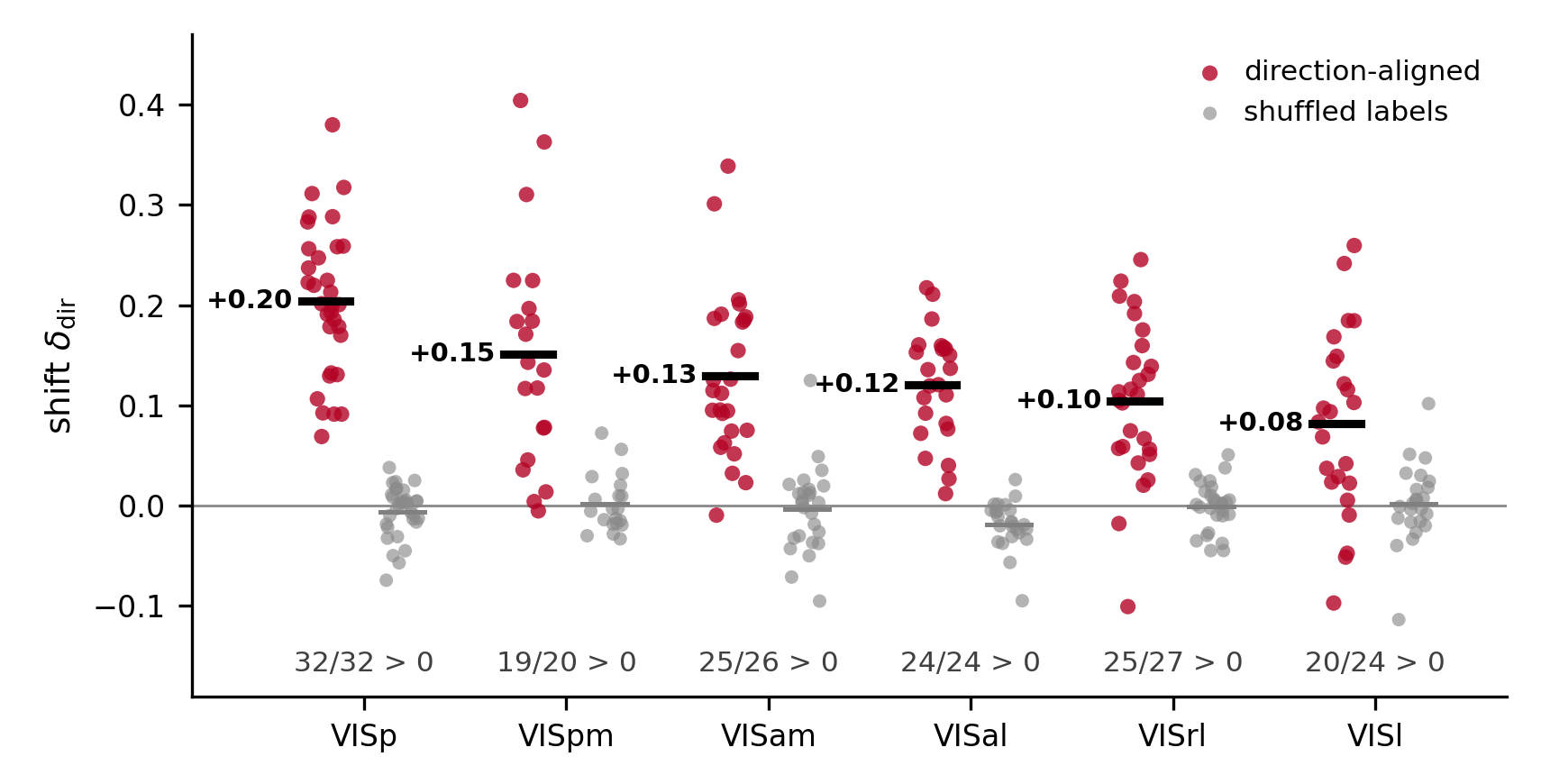}
\caption{Allen Neuropixels: one red point per session and area
(direction-aligned shift) for the six areas with at least 20
populations, gray points the shuffled-label controls, black bars the
area means. VISp is positive in all 32 sessions.}
\label{fig:allen}
\end{figure}

The replication spans animals and areas (Figure~\ref{fig:allen}): thirty-two
brain-observatory sessions, spike counts over each drifting-grating presentation (up to 600 presentations, eight directions), and the identical ladder, floor and shuffled-label control (three permutations per population), run per visual area with at least 50 units. Each direction class pools the five temporal frequencies, about 15 presentations each (about 6 in one VISp population), so the within-class part includes temporal-frequency tuning; \citet{ye2026} condition on both in the same release. Of the 167 session-area populations across 10 area labels, 158 have positive direction-aligned shifts (session identifiers, unit filters and windows are in the code repository).

Presentations that overlap an interval the release marks invalid for the population's probe (the probe sent no data), for all probes or for the stimulus are excluded, as are units on electrodes the release marks invalid and presentations in which a population whose rate predicts ten or more spikes fires none (one VISp population is silent for its first 2{,}135\,s); 11 of the 167 populations lose presentations (251 to 594 kept) and 7 lose 1 to 14 units.

The six areas with at least 20
session-area populations are tabulated below; the remaining four
(VIS, VISli, VISmma, VISmmp, totaling 14 populations) had fewer
sessions and are included in the 167-total and all reported
correlations but omitted from the table.
\begin{table}[h]
\centering
\small
\begin{tabular}{lcccc}
\toprule
Area & Populations & Mean $\delta_{\mathrm{dir}}$ & Positive & Mean $\bar\delta_{\mathrm{shuffled}}$ \\
\midrule
VISp & 32 & $+0.205$ & 32/32 & $-0.006$ \\
VISpm & 20 & $+0.151$ & 19/20 & $+0.002$ \\
VISam & 26 & $+0.129$ & 25/26 & $-0.004$ \\
VISal & 24 & $+0.120$ & 24/24 & $-0.019$ \\
VISrl & 27 & $+0.105$ & 25/27 & $-0.001$ \\
VISl & 24 & $+0.082$ & 20/24 & $+0.002$ \\
\bottomrule
\end{tabular}
\caption{Replication in a second laboratory and modality. Primary
visual cortex is positive in every session (one animal per session;
mean $+0.205$, SD $0.075$),
the same sign as and overlapping the calcium-imaging range. Per-session values,
per-area shuffle controls, and eigenspectra are in the code
repository.}
\end{table}
Random-axis controls on the same populations give mean
$\delta_{\mathrm{rand}} = -0.004$ (SD $0.030$); their spread is wider than imaging's: these populations hold 51 to 421 units, the imaging recordings 11{,}311 to 22{,}565 neurons. In the same release, the geometry of VISp's trial-to-trial covariance under static gratings differs between mice at least as much as the geometry of their mean responses \citep{duong2023}.

\paragraph{Permutation test per population.} Each population was
retested against 200 label permutations at fixed class sizes
(one-sided $p = (1 + \#\{\delta_{\mathrm{perm}} \geq
\delta\})/201$), with a 50-draw floor, whose shifts differ from the table's by a median absolute $0.008$. Populations at
$p \leq 0.05$: VISp 30/32, VISal 18/24, VISpm 15/20, VISrl 17/27,
VISam 15/26, VISl 12/24, VIS 2/8, the three smallest areas 4/6; 113 of
167 in total. Permutation SDs are $0.02$--$0.12$ at about 75 trials per
direction (31 in one population) and 51--421 units, so a single non-VISp population
resolves only large shifts; significance follows the same area gradient as the mean
shift. With the 50-draw floor 157 of 167 shifts are positive (158 with
the ten-draw floor of the table): three populations within $0.04$ of
zero change sign with the floor draw.

\begin{sloppypar}\paragraph{Shift and tuning amplitude.} The hierarchical model (response $\delta_{\mathrm{dir}}$; fixed effects quadrupole, dipole, cardinal fraction, area; random intercept per session) gives a quadrupole coefficient of $+0.92 \pm 0.17$ ($p = 5 \times 10^{-8}$, REML); an OLS with session-clustered robust standard errors agrees ($+0.98 \pm 0.25$, $p = 9 \times 10^{-5}$). Partial correlations of $\delta_{\mathrm{dir}}$ with the quadrupole amplitude are $+0.47$ given the dipole and $+0.39$ given population size and full-population PR; the even sector carries the association as a whole ($r(\delta, A_{\mathrm{even}}) = +0.42$, $r(\delta, A_{\mathrm{odd}}) = -0.03$, with $A_{\mathrm{even}} = \sqrt{b_2^2 + b_4^2}$, $A_{\mathrm{odd}} = |c_1|$), and the within-class PR of the first rung, where the ladder starts, is uncorrelated with the amplitude ($r = +0.02$) and enters the shift with the opposite sign (partial $r = -0.45$); this is the within-class geometry that the split of Section~\ref{sec:sign} resolves.\end{sloppypar}

The split-half cross-fit (harmonics from one half of each class's presentations, the shift from the other, and swapped) gives $r = +0.34$ and $+0.42$ against $+0.42$ on the same trials. Across Allen's populations the shift follows tuning amplitude; its partial correlation with the direction-selective fraction (median $0.21$) is $-0.12$ ($p = 0.12$) given amplitude, area and population size; splitting VISp's tuned units into direction-selectivity thirds is underpowered at this design (the imaging effect, resampled to 131 tuned units and 75 trials per class, passes 62\% of sessions; Allen passes 50 to 68\%). An orientation-binned variant (four classes) is positive in 150 of 167 populations (VISp mean $+0.206$, 30 of 32).

\paragraph{Temporal frequency separated.} Each direction class pools five temporal frequencies; their tuning is 19\% of the within-class variance in VISp (median; 10 to 31\%, from products across halves of direction-by-frequency cell means). Measuring each presentation against the leave-one-out mean of its own direction and temporal frequency, plus its direction mean, keeps the class means and removes that tuning: VISp's mean shift falls from $+0.204$ (this arm's floor draws; $+0.205$ in the table) to $+0.164$, positive in 30 of 32 mice (151 of 167 populations), with $p \leq 0.05$ in 27 of 32 against permutations of direction within each temporal frequency (22 at the minimum $1/201$). Units passing the Allen quality filters give 26 of 30. Section~\ref{sec:allen} opens on the pooled shifts of the table and uses this arm for the split, the state modes and the two terms. Directions are interleaved at random in every session (consecutive presentations share a direction at rate $0.125$, range $0.100$ to $0.152$, against $0.125$ at random), so the label permutation is the matched null.

\paragraph{Controls and the split.} Giving every class the same pooled fluctuations puts the within-class part within $0.03$ of zero in all 32 VISp populations and lowers the shift in 30, but leaves it positive in 27 of 32 (median $+0.078$), because Allen's class-mean part is positive; shuffling each unit within its class lowers the shift in all 32 (negative in 24).

The within-class part ($+0.102$, positive in 30 of 32; about 75 presentations per class against a within-class PR of 3 to 15, so the two parts are identified) exceeds what estimation alone puts into it: surrogates that keep each population's class means, sizes and variances with one covariance shape for every class give a median of $-0.011$ ($-0.063$ to $+0.022$), below the data in 31 of 32. The same surrogates reproduce the class-mean part (data minus surrogate, median $-0.012$): it is a property of the class means and the pooled fluctuations. It is positive in 27 of 32 VISp populations (158 of 167), and Eq.~\ref{eq:sign} gives its sign in 30 of 32 (162 of 167).

\paragraph{Why the class-mean part is positive.} Allen's fluctuations are no more concentrated than imaging's at matched size (the top 5\% of units hold $0.41$ to $0.55$ of the within-class variance, against $0.51$ to $0.56$); the class means' alignment $a$ differs ($0.24$ to $0.34$ against $1.08$ to $1.21$ for 64 to 256 units), and swapping $a$ alone between the two datasets swaps the sign at every size. Scaling each unit by its own within-class SD also turns the class-mean part negative (23 of 32), and Eq.~\ref{eq:sign} still gives its sign in 27 of 32.

\paragraph{The two-term account.} The two terms, computed from the counts, give the class-mean part at $r = 0.95$ (sign 161 of 167) and the shift at $r = 0.68$ ($0.84$ in VISp; sign 152 of 167, 31 of 32 in VISp; Figure~\ref{fig:twoterm}). The 15 misses are all populations whose shift is smaller than $0.10$ (median $|\delta| = 0.04$, against $0.12$ over the 167; one in VISp, seven in VISl), so every shift with $|\delta| \geq 0.10$ has the predicted sign (100 of 100); in all 15 the class-mean term has the measured sign, and the measured within-class part is negative in 13, in 12 of them where the mean field puts it positive. Their classes share more fluctuation (median $\sigma = 0.95$, against $0.87$ in the populations predicted right), and the mean field puts their within-class part near zero (median $+0.02$).

The overlap identity at the rungs' sample counts (Appendix~\ref{app:methods}'s finite-sample form, with the residualization's trace moments) gives the within-class part of all 167 populations at $r = 0.95$ (the mean field, $0.59$) and its sign in all 15 misses (mean $-0.083$ against the measured $-0.071$); with it the two terms give the shift's sign in 163 of 167 (31 of 32 in VISp) and track its size at $r = 0.94$. The class-to-class structure the mean field averages into one $\sigma$ carries the correction: at the same rungs the population identity already gives $-0.096$ over the misses, and finite sampling moves it by $+0.013$. Across the 167 populations there are 4.7 to 27.5 trials per fluctuation dimension, and the misses have no fewer than the hits (medians 11.5 and 10.2). The within-class term is the noisier at about 37 trials per class per half (split-half reliability $0.41$), so its size is not tested mouse by mouse.

\paragraph{State modes.} The largest within-class modes are shared and untuned. Removing the top three or five raises the median $a$ from $0.24$ to $0.79$ or $1.02$; in imaging, at 128 neurons, removing them lowers it from $1.12$ to $0.58$ and $0.44$: there the largest modes lie along the tuning, as gain fluctuations would. One of the top three tracks running speed at $|r| \geq 0.3$ in 29 of 32 mice (median of the largest $|r|$, $0.61$; speed alone explains a median 12\% of the trial-to-trial variance). Removing the three, estimated on one half of the trials and projected out of the within-class residuals of both halves with the class means kept fixed, makes the class-mean part negative in 31 of 32 (median $-0.27$; removing one, 19 of 32; five, 32 of 32) while the within-class part is positive in all 32 after removal (mean $+0.10$). In the same release a latent-gain model finds the shared gain covariance misaligned with the activity covariance \citep{thomas2026}, and across mouse cortex movement dominates single-trial variance \citep{musall2019}.

\begin{figure}[h]
\centering
\includegraphics[width=\linewidth]{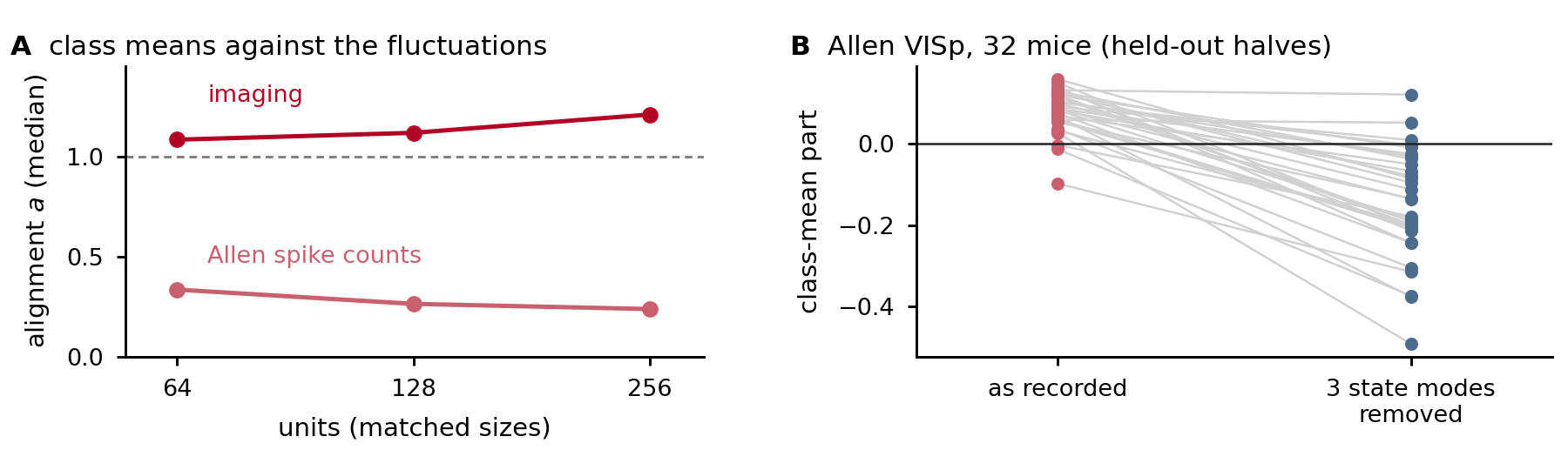}
\caption{Why Allen's class-mean part is positive, and what turns it negative. (\textbf{A})~Median alignment $a$ of the class means with the fluctuations at matched unit counts: above one in imaging, $0.24$ to $0.34$ in Allen, so Allen's means fill weak modes and raise PR (Eq.~\ref{eq:sign}). (\textbf{B})~Removing the three largest within-class modes, estimated on the first half of the trials and removed from the second, turns the class-mean part negative in 29 of 32 VISp mice (3 before).}
\label{fig:allen_state}
\end{figure}
\paragraph{State modes held out.} The halves alternate presentations within each direction-by-frequency cell, and each half is freed of its temporal-frequency tuning on its own trials. The top one, three or five within-class modes are estimated on one half and removed from the other, whose ladder and split are measured on that half alone. The class-mean part is negative in 3, 11, 29 and 30 of 32 mice at zero, one, three and five modes estimated on the first half (Figure~\ref{fig:allen_state}B), and in 2, 12, 30 and 31 the other way. At three modes its median moves from $+0.09$ and $+0.09$ to $-0.16$, while the within-class part moves from $+0.06$ to $+0.05$ and from $+0.08$ to $+0.05$ (median changes $-0.015$ and $-0.034$). These half-size within-class parts, at 35 and about 40 presentations per class, sit below the all-trial $+0.10$.

Regressing out running speed instead (loadings fitted on one half, applied with the other half's speed) makes the class-mean part negative in 7 of 32 each way: running is one part of these modes. In the 26 sessions with eye tracking, one of the three also tracks pupil area in 25 (median largest $|r|$, $0.61$), which indexes arousal beyond running (median largest partial $|r|$ given speed, $0.40$).

Pupil and running at lags of 0 to 1.5\,s, with pupil change, their squares and product and a running indicator, align with combinations of the three modes (median first and second canonical correlations $0.86$ and $0.52$); regressing them out does what removing the largest mode does (the class-mean part negative in 5 and 8 of 26, against 5 and 9), while removing two modes gives 16 and 19 and removing three 21 and 21 (on the presentations where every regressor is defined).

Running speed is associated with direction at $p < 0.05$ (2{,}000 permutations) in 2 of 32 sessions, one of them after Holm correction over the 32; its squared correlation ratio over the eight classes has a median of $0.013$, at its permutation expectation ($0.012$). With temporal frequency it is associated in 12 of 32 (median $0.008$).

\paragraph{Choosing an axis among candidates.} Four declared axes were screened in each VISp session: direction (eight classes, rungs as in Section~\ref{sec:results}), temporal frequency (five classes, rungs of one to five), running state (tertiles of each presentation's mean running speed, rungs of one to three) and random labels (eight classes). Each axis's shift is compared with 200 permutations of its labels within strata of the other variables (direction within frequency and running tertile, frequency within direction and tertile, running within direction and frequency, random labels within direction and frequency), so each null keeps the label's association with the variables of its strata. An axis is discovered on one half of a session at $p \leq 0.0125$ (0.05 shared over the four) and replicates when the other half gives $p \leq 0.05$.

Direction is discovered in 23 and 25 of 32 sessions (on one half or the other) and replicates in 22 and 24; temporal frequency is discovered in 5 and 6 (replicating in 4 and 4), running state in 3 and 4 (2 and 2), random labels in 1 and 0, replicating in neither. Across animals, with the mice split into two sets of 16 and each mouse scored by its $z$ averaged over the two halves, direction has the largest median on either set and again on the other, with a positive $z$ in all 16 mice of each set (medians $4.0$ and $2.9$); temporal frequency is positive in 13 and 14 of 16 (one-sided sign test $p = 0.011$ and $0.002$), and neither running state nor random labels passes that test on either set.

Picked by the largest $z$ alone, direction comes first on one half in 26 and 27 sessions and first again on the other, with $p \leq 0.05$, in 21 and 22; across all four axes and both halves, 19 of the 64 picks fail that test, and 10 of them rank second on the other half with $p \leq 0.05$: some mice carry a second organizing axis (in one, running state reaches $z = 7.0$ and $6.7$ beside direction's $7.9$ and $5.5$).

\paragraph{Floors under other covariance statistics.} The principal-angle dissimilarity between two classes, one minus the mean squared cosine between their top five residual directions, averaged over class pairs, has a label-blind floor of its own: the mean over 50 label permutations at the same class sizes is a median 69\% of the raw value across the 167 populations (14.8 to 133.5\% by population). The floor compresses the areas: raw medians are $0.35$ to $0.40$ in all six, corrected ones $0.18$ in VISp against $0.09$ to $0.12$ in the others.

The raw value does not follow the number of units (Spearman $0.10$); the sum of squared Frobenius distances between class covariances, estimated from unbiased traces, has a small floor (a median 14\%) and grows with the number of units. In imaging at 256 neurons, the median floor share across the eight recordings falls from 95\% at 8 trials per class to 85\% at 32 and 45\% at 256, and at 256 the floor-corrected value ranks the recordings as the mean-free shift does (Spearman $0.79$).

\paragraph{An analytic floor.} The floor is predictable without permutations, as Eq.~\ref{eq:null} predicts PR's floor: with the labels ignored two classes are independent samples, so an elliptical random-matrix model, the pooled spectrum with a trial gain of the measured kurtosis, predicts their expected overlap \citep{ledoit2011,couillet2022}. It matches the permutation floors to a median 2\% over the eight imaging recordings and six trial counts (the spectrum alone, 28\%) and 6\% over the 167 Allen populations, whose spike counts are less well described by a shared gain; there it puts the median floor share at $0.66$ against the measured $0.69$.

\paragraph{Macaque arrays.} Three Utah arrays in anesthetized macaque V1 \citep{smith2008,kohn2016crcns}: twelve drifting-grating directions, 200 trials each, spike counts from 0.28 to 1.28\,s after onset, units passing the SNR and rate criteria of \citet{cowley2016} (59 to 105). Directions were block-randomized \citep{smith2008}, so the label permutation is the matched null. The shift is $+0.146$, $-0.354$ and $+0.245$ ($p = 1/201$, $1.00$, $1/201$; 200 permutations); its within-class part $+0.242$, $+0.085$, $+0.173$ and its class-mean part $-0.096$, $-0.439$, $+0.072$. Shuffling each unit within its class turns the shift negative in all three ($-0.30$ to $-0.71$), and pooled fluctuations bring the within-class part to $-0.02$ to $0.00$.

The second array's class means carry $1.94$ times its fluctuations' variance and lie along them ($a = 0.8$ to $1.0$), in a preparation whose activity carries strong coordinated fluctuations under opioid anesthesia \citep{ecker2014}, so its class-mean part outweighs the within-class part, as the two terms predict ($-0.24$ against $-0.35$; for the other two, $+0.05$ against $+0.15$ and $+0.18$ against $+0.24$). Anatomical $3 \times 3$ electrode tiles span about an orientation hypercolumn at the arrays' 400\,$\mu$m pitch, too coarse to test the columnar prediction of Section~\ref{sec:discussion}.

\section{Declared-axis ladders on language models}
\label{app:llm}
The construction runs unchanged on a language model: prompts are the samples, a layer's hidden units the features, each
prompt read at its last token. It illustrates the instrument where no symmetry is known; the cortical claims rest on the
recordings alone. PR is a second moment of the covariance, so the argument that nearest-neighbour estimates of a layer's intrinsic dimension cannot track the dimension of its manifold \citep{schulte2026} does not bound it. A semantic axis has no natural order, so the reading that carries over between orders is the shift averaged
over random orders of the classes, against its own null: label permutations, each averaged over its own random orders.

\paragraph{One path inflates a shift.} A first battery, eight topics of eight prompts on Pythia-160m \citep{biderman2023},
accumulated the topics in alphabetical order. Along that path the shift is $+0.62$ at the embedding layer and $+0.15$ at the
last; averaged over 200 orders it is $+0.035$ to $+0.105$, and the alphabetical path sits at the 86th to 95th percentile of
the orders, where V1's angular order sits at the 15th to 90th (Appendix~\ref{app:orient}). The first topic on that path,
arithmetic, ends its prompts on two words, so at the embedding, which holds only the last token, its rung spans almost
nothing: that part of the shift is lexical.

\begin{figure}[h]
\centering
\includegraphics[width=\linewidth]{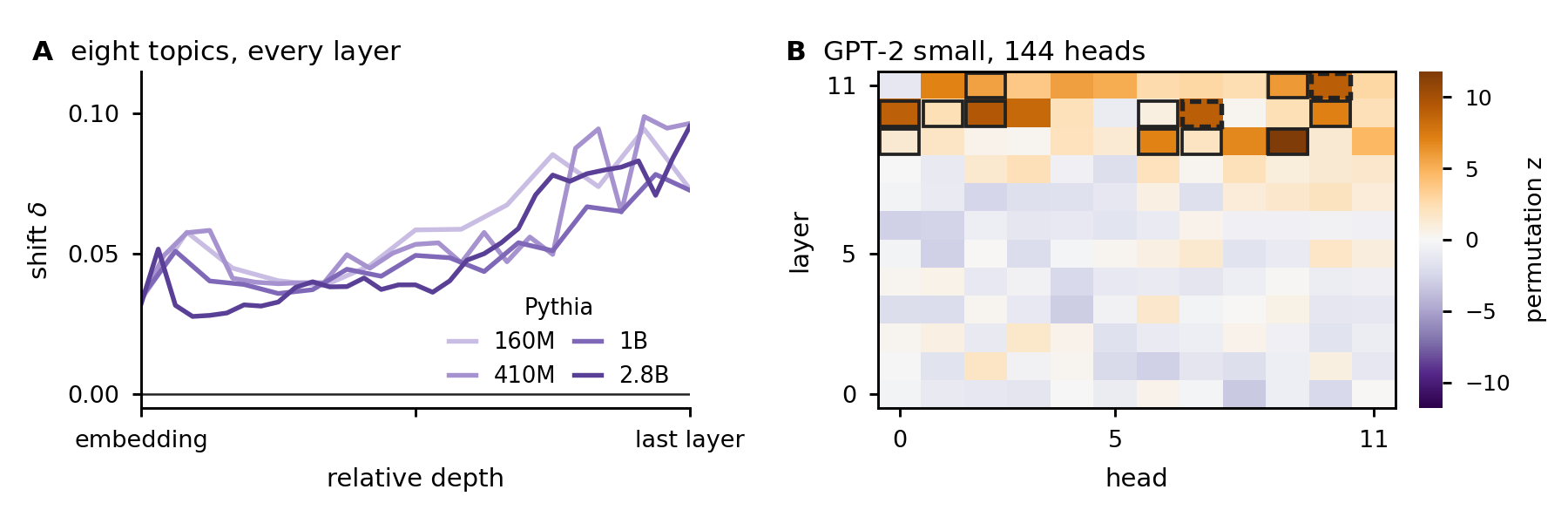}
\caption{Language models. (\textbf{A})~The content partition's order-averaged shift lies above its null at every layer of four Pythia models and grows with depth. (\textbf{B})~GPT-2 small: each attention head's permutation $z$ on the name axis (this run); solid boxes mark the eleven name-mover and backup heads found by path patching, dashed boxes the two negative name movers (AUC $0.92$ here, $0.93 \pm 0.01$ over ten seeds).}
\label{fig:lm}
\end{figure}
\paragraph{Declared partitions lift nearly every layer, and the content lift grows with depth.} Two partitions of 32 prompts per
class, eight topics (content: mathematics, geography, history, technology, law, medicine, philosophy, sports) and eight constructions, run on four Pythia models (160M to 2.8B parameters). Averaged over
orders, the content partition lifts the shift above its null at every layer of all four (Figure~\ref{fig:lm}A) (13 of 13, 25 of 25, 17 of 17 and 33
of 33 layers), the construction partition at all but a few layers (13/13, 21/25, 16/17, 33/33; the exceptions are the first three blocks and layer 21 of Pythia-410M and the first block of Pythia-1B).

The content lift grows
with depth, from $+0.03$ at the embedding to $+0.08$--$+0.10$ in the late layers, a rise beyond the null's at every size
($p = 0.005$ to $0.02$) whose sign holds in both halves of the prompts of Pythia-160m and 2.8B; the construction lift's rise is significant only at 2.8B ($p = 0.035$) and in neither half of its prompts. A declared
order's own depth profile does not reproduce: its trend changes sign between two samples of 16 prompts per class ($+0.10$
against $-0.06$ for the constructions on Pythia-160m), so a profile is reported averaged over orders.

Density-peak clusters of last-token states match a benchmark's subjects best in early layers when the prompt carries worked examples \citep{doimo2024}; the shift reads how a partition organizes the covariance, and its lift is largest late. Floors lie at $0.11$ to $0.80$ across layers,
near the crossover of Eq.~\ref{eq:null}, and shifts are compared across a model's own layers.

\paragraph{How much is the final token.} The number of distinct final tokens in a set of prompts moves the dimensionality estimated from their representations \citep{osman2026}, and here the number of distinct final tokens in a class tracks the class's own PR at the embedding (Spearman $0.82$ and
$0.84$ for the two partitions, Pythia-160m). Matching it across classes (25 and 28 distinct final tokens per class) in all four models, against random subsets
of the same size, removes the embedding's lift entirely ($-0.004$ to $0.000$, against $+0.017$ to $+0.026$), 65 to 100\% of the first three blocks' and 5 to 15\% of the late layers'. The lift that grows with depth survives: it stays above its null for the topics at 28 of 32 layers past the embedding of Pythia-2.8B (31 for
the size-matched subsets), 20 of 24 of Pythia-410M, 14 of 16 of Pythia-1B and 10 of 12 of Pythia-160m.

\paragraph{Benchmarks as axes.} A benchmark's categories are a declared axis ready to read, and each of the partitions below ships with \texttt{rung} as an example of what such an axis can carry. Partitions built from benchmarks at their largest balanced size (64 ARC questions \citep{clark2018arc}
and 64 HellaSwag contexts \citep{zellers2019hellaswag} per class, 26 TruthfulQA questions \citep{lin2022truthfulqa}) lift
Pythia-2.8B at 32, 19 and 31 of its 32 layers past the embedding; every TruthfulQA question ends on a question mark, so its lift
is not the final token, while ARC's classes are defined by keywords and so are lexical over the whole prompt. Whole-prompt wording can separate benchmark partitions with no shared computation behind them \citep{rahimi2026}, and matching final tokens controls the last word only. A partition by normative domain (the five ETHICS domains \citep{hendrycks2021ethics}, 64 scenarios each) is read more weakly, above its null at 3 of 12 layers of Pythia-160m and 12 of 32 of Pythia-2.8B, mostly late; its domains also differ in surface format.

\paragraph{A null must keep the nuisance.} Where the classes differ in a nuisance the permutation does not hold fixed, the
shift reads the nuisance. On BLiMP minimal pairs \citep{warstadt2020blimp}, eight classes of phenomenon by grammaticality, the
shift departs from the label permutation at nearly every layer, far below it, and swapping labels within pairs reproduces it at every layer: the classes are told apart by their carrier sentences. Built the other way, from minimal-pair contrasts of syntactic complexity \citep{baroni2026} (coordination against subordination, right-branching against centre-embedding, unambiguous against ambiguous attachment), and averaged over the orders that keep each pair together, the contrasts exceed the
within-pair swap at all 16 late layers of Pythia-2.8B ($+0.011$ to $+0.018$); a full random-order average dilutes such a
pairwise axis.

\paragraph{The shift beside a linear probe.} A probe reads where the class means sit; the shift reads how the classes spread.
With the seven directions the centred class means span projected out, so that a held-out probe falls to chance ($0.04$ to
$0.19$), the order-averaged shift stays above its own null (each null draw projecting out its own permuted subspace) at 28 of
33 layers for content and 33 of 33 for constructions in Pythia-2.8B (16 prompts per class): the declared classes differ in how their prompts vary as well as in where they sit, as the within-class part does in cortex. The converse holds on BLiMP: a
grammaticality probe trained on other pairs prefers the grammatical member of a held-out pair in 88 to 98\% of pairs at every
layer of Pythia-2.8B past the embedding, an offset between near-identical sentences that the shift, absorbed by the within-pair swap, cannot read.

\paragraph{A known circuit.} On GPT-2 small's indirect-object circuit \citep{wang2023ioi}, eight indirect-object names with 16 prompts each in balanced templates (the indirect object's logit exceeds the subject's in all 128) give each attention head's output an order-averaged shift along the name axis. Scored by its permutation $z$ (Figure~\ref{fig:lm}B), the shift ranks the eleven name-mover and backup heads that path patching found high among all 144 heads (AUC $0.92$ in this run and $0.93 \pm 0.01$ over ten seeds, mean $\pm$ SD; the three name movers alone $0.98$), with no classifier trained; a cross-validated probe's accuracy ranks them at $0.95$ and the class-mean variance fraction at $0.96$.

The ranking takes one forward pass per prompt and no patching; for shortlisting heads before path patching the class-mean share is the stronger statistic here, and the shift adds the permutation null and the split into means and fluctuations. The ranking reads how the axis organizes each head's write; which heads the output uses takes an intervention. The two negative name movers, which write the name with the opposite sign and were not in the positive set, rank third and fourth.

\section{Ground-truth symmetry test in an equivariant network}
\label{app:cnn}
Architectures with built-in symmetry supply what cortex cannot: a
representation whose symmetry content is known by construction, so
the rotation ladder has a known null expectation. Every MNIST image
receives a
random rotation from $C_4 = \{0^\circ, 90^\circ, 180^\circ,
270^\circ\}$ (exact \texttt{rot90}, no interpolation). Two digit
classifiers share an identical three-layer convolutional trunk: a
plain CNN, and an orbit-weight-sharing equivariant network (a $C_4$ group convolution; \citealp{cohen2016}) that
applies the same trunk to all four rotations of the input, yielding
an exactly $C_4$-equivariant feature stack whose mean over the orbit
is invariant by construction.

Rotation ladders accumulate the four
rotation classes in cyclic order against matched random floors;
digit ladders accumulate eight digit classes. The construction removes
rotation structure from the invariant layers, while the plain network and
the pre-pooling stack see the rotated strokes; the five-seed means show
rotation structure in both and none in the invariant layers, digit structure
at the last layer, and shuffled-label controls (one seed) near zero:
\begin{table}[H]
\centering
\small
\begin{tabular}{lcc}
\toprule
Features & $\delta_{\mathrm{rot}}$ (conv1/2/3) & $\delta_{\mathrm{digit}}$ (conv3) \\
\midrule
Plain CNN & $+0.015$ / $+0.092$ / $+0.068$ & $+0.035 \pm 0.016$ \\
Equivariant, pre-pool & $+0.014$ / $+0.093$ / $+0.096$ & $+0.041 \pm 0.019$ \\
Equivariant, invariant & $+0.000$ / $-0.000$ / $-0.009$ & $+0.047 \pm 0.011$ \\
\bottomrule
\end{tabular}
\caption{Five-seed means. Per-layer $\delta_{\mathrm{rot}}$ SDs:
conv1 $0.004$--$0.006$, conv2 $0.022$--$0.029$, conv3
$0.050$--$0.060$; invariant layers $\leq 0.014$; shuffled-label
controls $|\bar\delta| \leq 0.011$ (one seed). Because the pooled representation
is exactly invariant by construction, the rotation ladder has a
known null expectation, $\delta_{\mathrm{rot}} = 0$; the measured
values verify it while the digit axis stays positive on the same
features.}
\end{table}

The sector content is also resolved. The fraction of feature variance
carried by the rotation-class means grows with depth in both networks
(plain $0.006 \to 0.088$; pre-pool $0.008 \to 0.095$) and is
$97$--$99\%$ frequency-2 in the $C_4$ Fourier decomposition:
handwritten strokes have orientation but weak direction, so the rotation-tuned variance is mostly $\pi$-periodic, mirroring the even-led class means of the full-field V1 recordings (Section~\ref{sec:symmetry}). On the invariant layers the tuned fraction collapses
to $0.0005$ at every depth, the sector-resolved form of the null.
The sector content is stimulus-inherited (raw pixels are already frequency-2 dominant) and the probe reads it wherever the representation retains it; the invariance that removes it is architectural.

The fluctuation overlaps name the accumulation order here as in cortex (Appendix~\ref{app:order}). In every layer that carries rotation structure (the plain network and the pre-pool stack), and in every seed, opposite rotations share more of their fluctuation shape than neighbours ($0.99$ to $1.00$ against $0.62$ to $0.997$; odd harmonic power below 3\% of the even). Adding the opposite rotation first climbs faster than the cyclic order at each such layer of both networks (five-seed means $+0.002$ to $+0.012$; five of five seeds at conv1--2, four of five at conv3), and at conv3 the within-class part carries the effect.

\section{Methods and the covariance split}
\label{app:methods}
\paragraph{Estimator and inference.} All comparisons use the same centered Gram-trace estimator,
$\PR = (\mathrm{Tr}\,G)^2/\mathrm{Tr}(G^2)$ with each subset centered
across the elements it samples: each unit across the presentations of a stimulus rung, each presentation across the units of a neuron rung; a declared ladder and its floor share the estimator, and the shift depends on that choice. On GT3 the direction-aligned $\delta$ is
positive under both the centered ($+0.24$) and uncentered ($+0.46$)
estimators; centering is the conservative choice used throughout.

Observed accumulation ladders are
nested and deterministic given the class labels; floors use ten independent
random-subset ladders, each rung a fresh draw, and $\tfloor$ is the slope of the mean null
log-PR. The random-stimulus ladders of Appendix~\ref{app:random} average log PR over five draws per rung, as their floors do over ten ladders, so the two arms share one expectation; the spatial ladders of Appendix~\ref{app:neuron} average PR over their 50 anchors in both arms. Rungs with fewer than ten observations are skipped.

Inference uses 200 label permutations per cell, one-sided $p = (1 + \#\{\delta_{\mathrm{perm}} \geq \delta\})/201$; the shuffled-label means in the tables (three to ten permutations) are descriptive. With pseudo-labels that carry no stimulus information the test rejects at its nominal rate (2 and 7 of 100 at $\alpha = 0.05$, two recordings); labels constant within blocks of 50 presentations would be rejected 28 to 52\% of the time, because slow drift reads as structure, but the directions here are interleaved in time (consecutive presentations share a class 11.7 to 12.0\% of the time, against 12.5\% at random). Permuting directions only within four-minute epochs that hold every direction, which keeps slow drift and state, leaves the shift above all 200 such permutations in all eight recordings (null means $-0.007$ to $+0.007$).

The
resampling intervals draw 80\% of each class's trials without
replacement at full neuron count (100 replicates, percentile
intervals); the floor ladder is redrawn inside every replicate, so the
intervals include floor-estimation noise (point estimates use the
ten-draw mean floor; on the random ladders the floor's slope
dispersion is $0.01$--$0.03$ in the grating recordings). The
intervals describe subsampling stability within a recording (no coverage is claimed); animal-level inference rests on the 32
single-animal sessions of Section~\ref{sec:allen}.

\paragraph{Harmonics of the class means.} With eight classes the five cosines $\ell = 0$ to $4$ are the complete symmetric basis of the class-mean profile, so its sector split is exact and every shorter fit truncates it. The separation average $C(\Delta\phi)$ explains $79$--$96\%$ of the off-diagonal variance of the full $8 \times 8$ class-pair correlation (GT3 $96\%$; largest residual $0.17$, low-contrast GT1). The departures reproduce across disjoint halves of the presentations (split-half $r \geq 0.99$ for the off-diagonal entries; circulant share within $0.02$ between halves), so they are a reproducible anisotropy of the code across directions.

The multipole coefficients are the exact discrete Fourier transform of the eight-direction class-mean correlation profile (GT3: the fit $a + b\cos 2\Delta\phi + c\cos\Delta\phi$ gives $a = 0.39$, $b = 0.33 \pm 0.05$, $c = 0.14 \pm 0.05$, $R^2 = 0.90$; adding $\cos 4\Delta\phi$ raises $R^2$ to $0.97$ with $b_4 = 0.07$; $c_3 = 0.064$); the quadrupole
exceeds the dipole in all five full-field grating recordings
($b/c = 1.19$--$2.40$) and not in the three localized ones
($0.50$--$0.78$). The static sessions span too narrow a range for an
eight-class profile.

\paragraph{Direction-selective thirds.} The direction-selective thirds of
Section~\ref{sec:symmetry} take each neuron's direction and orientation selectivity as vector strengths of its eight class means (minimum subtracted),
keep neurons tuned at $p < 0.01$ (F-test across classes), and run the
eight-class direction ladder against ten-draw floors on three subsets
of equal size (a third of the tuned neurons): the most and least
direction-selective thirds and a random third.

The most selective
third is dipole-dominant and the least quadrupole-dominant in all eight recordings (profiles of class means standardized per neuron); selected on alternate halves of each class's trials and measured on the other half, the most selective third stays above the least selective in eight of eight recordings, both ways round; the direction-selective fraction is the share of tuned neurons with selectivity above $0.3$.

After half the circle an exactly even code has every class mean it will have: there the least selective third has closed two thirds or more of its first-rung deficit to the floor, while the most selective third still holds $37\%$ to $102\%$ of its own. A random third, which carries both sectors, exceeds the most selective third in two full-field drifting recordings ($0.41$ against $0.35$, $0.42$ against $0.36$): the shift reads both sectors, and the direction-selective neurons carry its odd part.

\paragraph{Where the second half's new directions come from.} The late gain is the ladder's slope over its 4-, 6- and 8-class rungs. Giving each pair of opposite classes one shared mean keeps $-2$ to $36\%$ of it in the full-field drifting recordings, whose odd sector sits in the class means, and 85 to 88\% in the localized ones; giving opposite classes shared fluctuations instead keeps 67 to 88\% in the drifting recordings and 16 to 22\% in the localized ones; at low contrast each removes about half.

Across the eight recordings the shift's correlation with the direction-selective fraction ($r = +0.87$) is the localized-against-full-field contrast: with mouse and protocol means removed it is $+0.06$ (exact restricted $p = 0.92$). The fluctuation geometry survives the same removal: the more direction-selective the population, the less its opposite classes share fluctuation directions ($r = -0.84$ with mouse and protocol removed; negative within every protocol and with any mouse left out).

\paragraph{Within-class and class-mean parts of the shift.}
For a subset with class labels, write its centered covariance as
$C = W + B$, with $W$ the pooled within-class covariance (trials minus
their class mean) and $B$ the class-size-weighted covariance of the
class means about the subset mean, both normalized by the subset's
trial count, so that $C = W + B$ exactly (the law of total
covariance; the pair behind neural collapse in trained classifiers, \citealp{papyan2020,chou2026}). With
$w_1 = \mathrm{Tr}\,W$, $w_2 = \mathrm{Tr}\,W^2$, $b_1 = \mathrm{Tr}\,B$,
$b_2 = \mathrm{Tr}\,B^2$ and $x = \mathrm{Tr}\,WB$,
\[
  \log \PR(C) = \underbrace{\log\frac{w_1^2}{w_2}}_{\text{within-class}}
  + \underbrace{\log\frac{(1 + b_1/w_1)^2}{1 + (b_2 + 2x)/w_2}}_{\text{class means}}
\]
exactly.

Applying the split to every rung of the declared ladder and of
its floor (whose random subsets carry their own class labels) and
taking slopes splits $\delta$ into two additive parts. The split is a convention: the class-mean part is the increment of $\log\PR$ over the within-class reference, the overlap $\mathrm{Tr}\,WB$ included.

\paragraph{Finite-sample bias of the split.} When every class shares one within-class covariance the within-class part vanishes in the population limit; at finite samples each rung subtracts the mean of every class it holds, so a declared rung with few classes and its floor keep different residual degrees of freedom. For eight identical Gaussian classes in the tested designs the bias is $-0.054$, $-0.012$, $-0.003$ at $30$, $100$, $300$ trials per class ($20$, $50$, $100$ isotropic dimensions) and $+0.005 \pm 0.004$, $-0.002 \pm 0.004$ at $300$ and $500$ trials per class ($300$ and $200$ power-law dimensions), against $390$--$681$ trials per class and a within-class part of $+0.29$ to $+0.64$ in the recordings. The power-law designs have effective dimension $24$ and $21$, within the range of a single class's PR in the recordings (18 to 38 in drifting gratings, 19 to 59 localized, 58 to 91 at low contrast).

On the recordings themselves, surrogates that keep every class mean and give every class one fluctuation shape put the within-class part at $-0.002$ to $-0.030$. For identical classes estimation only moves weight between the two parts: eight identical isotropic classes at 25 per class give a within-class part of $-0.15$ and a class-mean part of $+0.15$. Classes that differ add the floor's sampling term (\emph{Mean field and harmonics} below), and the per-class count is reported against $\PR(W)$ beside every split.
Table~\ref{tab:split} gives them for the direction ladder of
Section~\ref{sec:results} (all neurons; independent floor draws, whose
$\delta$ matches Figure~\ref{fig:ladder}'s within
$0.014$) and for the order effect of Appendix~\ref{app:order}
(paired minus sequential; within $0.019$ of the values in
Appendix~\ref{app:order}).
\begin{table}[h]
\centering
\small
\begin{tabular}{lcccccc}
\toprule
 & \multicolumn{3}{c}{direction ladder} & \multicolumn{3}{c}{order effect} \\
\cmidrule(lr){2-4}\cmidrule(lr){5-7}
Recording & $\delta$ & within-class & class means & total & within-class & class means \\
\midrule
Drifting GT1 & $+0.351$ & $+0.578$ & $-0.227$ & $-0.031$ & $-0.171$ & $+0.140$ \\
Drifting GT2 & $+0.292$ & $+0.642$ & $-0.350$ & $-0.066$ & $-0.128$ & $+0.062$ \\
Drifting GT3 & $+0.233$ & $+0.362$ & $-0.129$ & $-0.117$ & $-0.046$ & $-0.072$ \\
Local GT1 & $+0.455$ & $+0.559$ & $-0.104$ & $-0.131$ & $-0.191$ & $+0.060$ \\
Local GT2 & $+0.393$ & $+0.524$ & $-0.131$ & $-0.067$ & $-0.149$ & $+0.082$ \\
Local GT3 & $+0.460$ & $+0.574$ & $-0.113$ & $-0.021$ & $-0.104$ & $+0.083$ \\
Low-contrast GT1 & $+0.209$ & $+0.306$ & $-0.098$ & $-0.045$ & $-0.042$ & $-0.003$ \\
Low-contrast GT2 & $+0.217$ & $+0.295$ & $-0.078$ & $-0.035$ & $-0.022$ & $-0.013$ \\
\bottomrule
\end{tabular}
\caption{The split of the shift. The within-class part exceeds $\delta$
and the class-mean part is negative in all eight recordings. A
positive within-class part means the pooled within-class covariance
gains dimensions as classes are added; in the population limit that requires the within-class
covariance to differ between classes, since with one shared within-class covariance the part vanishes, and estimating each class mean biases it by at most $0.03$ at these trial counts. The class-mean part is negative when the class means add
a few large directions, which concentrates variance. The order
effect (declared origin) has its within-class part negative in all eight and its class-mean
part favouring the paired order in five; Appendix~\ref{app:order} gives the other origins.}
\label{tab:split}
\end{table}

\paragraph{Tuning inside a bin.} Directions are continuous, so a $45^\circ$ class also holds tuning inside its bin. Replacing each presentation by its residual from the leave-one-out mean of its $7.5^\circ$ sub-bin, plus its class mean, removes that tuning and keeps the class means exactly (leave-one-out residuals sum to zero in each sub-bin, and the sub-bins tile each class): the shift stays positive in eight of eight and its within-class part positive in eight of eight, and the order effect's within-class part stays negative in eight of eight.

In the five localized and low-contrast recordings, where tuning inside a bin carries $4$--$6\%$ of the within-class variance, nothing changes ($|\Delta\delta| \leq 0.03$). In the three full-field drifting recordings it carries $11$--$17\%$, and removing it lowers the shift ($+0.351 \to +0.301$, $+0.233 \to +0.163$, $+0.292 \to +0.032$): there, part of the class-dependent covariance is the tuning slope across each bin, which adds directions as classes enter by the same geometry as the fluctuations do.

In Drifting GT2 this tuning carries most of the shift from every origin (removing it lowers $\delta$ by $0.25$ averaged over the eight origins, against $0.10$ in GT1 and $0.04$ in GT3): its shift is the smallest margin between a large within-class part ($+0.64$) and the most negative class-mean part ($-0.35$), its class means being the strongest of the eight ($\mathrm{Tr}\,B/\mathrm{Tr}\,W = 0.28$).

\paragraph{What carries the shift.} Shuffling each neuron's trials within each class, independently per neuron, keeps every class mean and every neuron's per-class variance and removes the correlations between neurons, including the covariance that tuning inside a $45^\circ$ bin creates. On the direction ladder of Table~\ref{tab:split} (same seed) $\delta$ falls from between $+0.21$ and $+0.46$ to between $-0.33$ and $+0.07$, negative in seven of eight recordings. The within-class term keeps a positive share under the shuffle ($+0.08$ to $+0.39$), since a class's variance still concentrates on its preferred neurons, while the class-mean term turns more negative: the two terms are a covariance accounting, and the shift's sign needs the correlated fluctuations in seven of eight, as in the calibrated model below, where isotropic within-class variability also reverses the sign ($-0.57$).

Two controls run on the tuning-free residuals of the preceding paragraph, each reported as a mean of three seeds. The same shuffle gives $-0.44$ to $+0.05$, below the residual shift in eight of eight and negative in seven. Redistributing whole residual vectors across all presentations, which keeps the correlations between neurons and removes their dependence on the class, puts the within-class part within $0.03$ of zero and the shift at $-0.41$ to $-0.07$, negative in eight of eight: correlated fluctuations that differ between classes carry the shift.

On the original responses the positive shift also survives per-neuron z-scoring, the correlation-matrix PR ($+0.10$ to $+0.26$), and a square-root transform of the responses ($+0.01$ to $+0.27$), in eight of eight. The within-class part vanishes when the classes share one covariance, the hypothesis Box's $M$ tests \citep{box1949}, and also when their covariances differ only in scale: it reads differences in the shape of the class covariances.

\paragraph{Correlations and information.} Removing the correlations changes $\delta$; whether the same correlations limit information is a separate question. That limit is set by the noise along the tuning derivative \citep{morenobote2014}, a small part of the correlated fluctuations in mouse V1 \citep{rumyantsev2020}, where information about direction grows sublinearly with population size \citep{kafashan2021}; covariance-geometry nulls separate a shuffle effect from discriminative optimality \citep{livezey2025}.

\paragraph{Two kinds of difference between the classes' fluctuations.} On the tuning-free residuals of all neurons, surrogates that keep every class mean and every neuron's per-class SD but give all classes one correlation matrix (recovered at $r = 0.997$ to $0.998$) carry 41 to 57\% of the within-class part and the order effect's sign in all eight recordings; identical classes give $-0.002$ to $-0.030$. The per-class variances follow a Fano-like law, variance linear in the class mean ($\sigma^2_{ic} = v_i + s^2 m_{ic}$; fit $r = 0.74$ to $0.86$), which reproduces those profiles' within-class part in seven of eight and their order part in eight of eight; a pure power law overshoots and a gain on the squared tuned response reproduces neither. The rest of the within-class part is correlation structure that changes with direction (in Allen's spike counts the shared-correlation surrogate falls $0.07$ short, median).

The neighbour-over-opposite overlap of Figure~\ref{fig:multipole}C survives in the correlation frame and with neuron variances permuted in all eight recordings (it roughly halves), so it is not a by-product of which neurons fluctuate most.

\paragraph{Decoders.} Direction was decoded from 256 randomly drawn neurons per recording by five-fold cross-validated multinomial logistic regression (accuracy = fraction of held-out presentations assigned their class; chance $0.125$). Its accuracy, $0.67$ to $0.92$, is higher where the class-mean part of Table~\ref{tab:split} is more negative (Spearman $-0.98$ over the eight recordings; with each protocol's mean removed, Pearson $-0.71$), and ranks the recordings unlike the shift ($+0.24$). The class means' strength $\beta = \mathrm{Tr}\,B/\mathrm{Tr}\,W$ is what decoding and the class-mean part share: at the decoder's 256 neurons (ten draws) it ranks the recordings exactly as accuracy does (Spearman $1.00$; $0.93$ at all neurons) and in reverse order to the class-mean part ($-0.98$).

A linear (LDA) and a quadratic decoder (QDA, one covariance per class, regularization chosen by inner cross-validation), in 16 standardized principal components of 256 neurons (five draws), give LDA accuracies of $0.42$ to $0.92$ across the recordings and their most and least direction-selective thirds. QDA gains at most $0.043$ over LDA, and its gain does not follow the within-class part, across recordings (Spearman $+0.31$) or between the thirds of one recording (larger in the most selective third in two and three of eight, cross-fitted both ways, where the shift is larger in eight of eight). In the same release's static-grating recordings the leading noise components are not aligned enough with the signal to bound information \citep{moosavi2026}.

\paragraph{Class means removed.} The seven directions spanned by a set of trials' centred class means (all neurons) are projected out of the responses, which makes the class means equal. For the decoders, each cross-validation fold estimates the seven directions on its training folds and removes them from training and test folds alike; on halves that alternate within each class, the logistic decoder (256 and 1{,}024 neurons) and LDA then score $0.129$ to $0.134$, the share of the largest class in each half, which checks the removal. Under the same removal QDA still reads direction, at $0.21$ to $0.67$ in all 16 halves (chance $0.125$).

Inference uses halves split without the labels, by odd and even presentations and by early and late ones (32 recording-by-half cells). There QDA, as above, still reads direction at $0.47$ to $0.69$ in the drifting and localized recordings and $0.21$ to $0.23$ at low contrast, above all 50 permutations of its whole procedure in every cell (null maxima $0.14$ to $0.15$), so the bound on its gain with the means present is not a limit of its capacity.

With each half's own class-mean subspace removed, the shift is $+0.04$ to $+0.51$, above all 200 permutations that each remove their own permuted subspace in 31 of the 32 cells and above 196 in the last (low-contrast GT1, early half); permutations confined to contiguous epochs of 48 presentations, which keep slow drift, give the same verdict ($p \leq 0.02$ in all 32). Its class-mean part is $+0.002$ to $+0.006$: the projected shift is almost entirely its within-class part. It also stays above a null that gives every class the same pooled fluctuations, a test of class-dependent covariance alone (all 100 draws, in 16 of 16 odd and even halves), and, with the tuning inside each $7.5^\circ$ sub-bin removed first, above all 100 of its label permutations in 16 of 16, keeping 67 to 103\% of itself.

On alternating halves, a subspace estimated on the other half leaves the shift above its null as well ($+0.07$ to $+0.50$, $p \leq 0.01$ in 16 of 16), but it removes the means only up to its own estimation noise, a fixed offset per class that a linear decoder reads ($0.27$ to $0.33$ at 256 neurons). The removal keeps $0.88$ to $0.95$ of the within-class part in the localized recordings, $0.56$ to $0.79$ in the full-field drifting ones and $0.36$ to $0.61$ at low contrast. Removing the other half's class-mean directions instead costs as much as removing the half's own ($r = 0.97$), while seven random directions drawn from the fluctuations or from permuted labels cost at most $0.16$ on average over 20 draws: the lost part is fluctuation structure along the class means themselves. Appendix~\ref{app:llm} runs the same design on a language model.

\paragraph{The within-class part as covariance overlap.} For the classes $A$ a rung holds, with class covariances $\Sigma_c$ and weights $p_c$, $W_A = \sum_c p_c\Sigma_c$ and, exactly,
\[
\frac{1}{\PR(W_A)} = \sum_{c,d\in A}\alpha_c\alpha_d K_{cd}, \qquad \alpha_c = \frac{p_c\operatorname{Tr}\Sigma_c}{\sum_d p_d\operatorname{Tr}\Sigma_d}, \qquad K_{cd} = \frac{\operatorname{Tr}(\Sigma_c\Sigma_d)}{\operatorname{Tr}\Sigma_c\operatorname{Tr}\Sigma_d}.
\]
The diagonal $K_{cc} = 1/\PR(\Sigma_c)$ is each class's spectral concentration; the off-diagonal terms are the variance-weighted overlap of two classes' fluctuations. Identical covariances keep $\PR(W_A)$ constant as classes enter; classes of equal variance weight $\alpha_c$ and equal PR whose fluctuations occupy disjoint directions add their dimensions; an order changes which overlaps each rung pools. $K_{cd}$ is unchanged when a covariance is rescaled, which is why, in the population limit, differences in scale alone leave the within-class part at zero. Class-conditional second-order geometry also drives supervised dimensionality reduction \citep{herreraesposito2025}; here the overlaps set how pooled within-class PR grows.

\paragraph{Mean field and harmonics.} If the classes carry equal weight and every pair of distinct classes overlaps by the same fraction $\sigma$ of a class's overlap with itself, $k$ classes give $\PR(W_k) = k\,\PR(\Sigma)/[1 + (k-1)\sigma]$: $\sigma = 1$ keeps the PR constant as classes enter, $\sigma = 0$ multiplies it by $k$. Measured with split halves, so that the diagonal carries no sampling term, $\sigma$ is $0.15$--$0.25$ in the drifting and localized recordings and $0.42$--$0.50$ at low contrast, and $\ln[8/(1+7\sigma)]/\ln 8$, the mean-field growth over the ladder's eight classes, reconstructs the within-class part of Table~\ref{tab:split} across the eight recordings ($r = 0.92$, mean error $0.054$, no fitted parameter, the same data); with $\sigma$ from one half of the trials and the part from the other, $r = 0.93$, the prediction overshooting by $0.12$ on average.

Finite sampling is the main source of the overshoot. The floor's first rung estimates the pooled fluctuations, of dimension $m\,\PR(\Sigma)$ with $m = K/[1+(K-1)\sigma]$, from as many samples as the declared first rung spends on one class, so the floor climbs as its rungs fill. With Eq.~\ref{eq:null} in both ladders, comparing the first rung with the full set, the within-class part at $n$ samples per class sits below the mean-field law by about the sampling term $s = (m-1)\,\PR(\Sigma)/(n\ln K)$, which halving the trials doubles (fitted over all rungs, its coefficient is 3 to 7\% smaller).

The Wishart moments behind Eq.~\ref{eq:null}, with each class's measured fourth moment in place of its Gaussian value $t_c^2 + 2T_{cc}$, give a finite-sample approximation to the identity: with $n_c$ samples of class $c$ in the rung, $\nu_c = n_c - 1$, $t_c = \operatorname{Tr}\Sigma_c$, $T_{cd} = \operatorname{Tr}\Sigma_c\Sigma_d$ and $\mu_{4,c}$ the class's mean fourth power of a fluctuation's norm,
\[
\PR_{\mathrm{fin}}(W_A) = \frac{\big(\sum_{c} \nu_c t_c\big)^2}{\sum_{c,d} \nu_c\nu_d T_{cd} + \sum_c \nu_c\,(\mu_{4,c} - T_{cc})}.
\]

With nothing fitted, this form gives the within-class part of the eight recordings (mean error $-0.002$; $-0.003$ with covariances and part from different trial halves) and of all 72 lattice and spin-wave ensembles ($r = 0.99$; Appendix~\ref{app:lattice}), where the population identity alone gives $r = 0.80$. In the recordings finite sampling lowers the part by $0.07$ to $0.13$ and the overlaps' dependence on separation adds $-0.08$ to $+0.11$; their means partly offset ($-0.10$ against $+0.05$) and the finite-sample correction varies little across recordings, so the mean field tracks the part with an offset. For eight equal Gaussian classes of $\PR(\Sigma) = 30$ the finite-sample form puts the within-class part within $0.05$ of the population identity's value on the same rungs at about 20, 8 and 3 times $\PR(\Sigma)$ samples per class for $\sigma = 0.2$, $0.44$ and $0.7$, at higher ratios for classes of lower PR; Figure~\ref{fig:twoterm} marks the lattice ensembles whose $s$ exceeds $0.05$.

\begin{sloppypar}Beyond mean field, equal class weights ($\alpha_c = 1/k$) and a kernel that depends only on the separation of two classes make the identity a structure factor, $1/\PR(W_A) = k^{-2}\sum_\ell \kappa_\ell\,|\rho_A(\ell)|^2$, with $\kappa_\ell$ the angular harmonics of $K$ and $\rho_A(\ell) = \sum_{c\in A} e^{i\ell\phi_c}$ over the classes a rung holds. An antipodally paired set has $\rho_A(\ell) = 0$ at every odd $\ell$, so the paired order reads only the even harmonics of the fluctuation overlaps, while an adjacent arc weights the low harmonics of both parities. This separation-only kernel gives the within-class order effect averaged over origins to $0.011$ in all eight recordings (Appendix~\ref{app:order}).\end{sloppypar}

\paragraph{What sets the sign of the class-mean part.} At the full stimulus set the class-mean term is $c = \log[(1+\beta)^2/(1+(b_2+2x)/w_2)]$ with $\beta = b_1/w_1$. Since $b_2/w_2 = \beta^2\,\PR(W)/\PR(B)$ and $2x/w_2 = 2a\beta$ with $a = \PR(W)\,x/(w_1 b_1)$, exactly
\[
  c = \log\frac{(1+\beta)^2}{1 + \beta^2\,\PR(W)/\PR(B) + 2a\beta}, \qquad c < 0 \iff \frac{\PR(W)}{\PR(B)} > 1 + \frac{2(1-a)}{\beta},
\]
which is Eq.~\ref{eq:sign}. The alignment $a$ is the fluctuation variance along the class means, $x/b_1$, relative to the variance-weighted mean eigenvalue of $W$, $w_2/w_1$: $a = 0$ when the means are orthogonal to the fluctuations, $a = 1$ when they lie along directions of typical weight. With $a = 0$ the rule is a sufficient threshold, $1 + 2/\beta$.

In the eight recordings $\PR(W)$ is 73 to 173 against $\PR(B)$ of 4.3 to 5.8, and the class means lie in the fluctuation subspace at 77 to 154 times chance (against the same class means with their neurons permuted) with $a = 0.96$ to $2.1$, as gain fluctuations along the tuning would place them \citep{lin2015}. The alignment weighs the class means against the whole fluctuation spectrum: the difference between neighbouring class means holds $0.2$ to $3.2\%$ of the largest fluctuation mode's direction, less than that mode's $3.5$ to $7.6\%$ share of the fluctuation variance in all eight recordings, so it lies nearly perpendicular to that mode, as in \citet{rumyantsev2020}, while its own alignment is $0.8$ to $2.1$.

Subsampled to 64 to 512 neurons the fluctuations fall below the $a = 0$ threshold (ratio 3 to 11 against 8 to 57) and the term stays negative, because $a$ stays near $1.1$: the full rule gives its sign in all 56 recording-by-size cells (64 neurons to all), the $a = 0$ threshold in 15. In the full-population imaging ladders whose first rung holds one class, which holds no class-mean variance while the floor draws across all classes, the class-mean part takes $c$'s sign: across 1{,}536 such ladders (48 class orders, four rung schedules, eight recordings) the sign never departs from $c$'s, while 117 of the 384 ladders that start at two classes do, in every recording (11 to 20 of 48 each): a two-class first rung already holds class-mean variance, and which pair enters first can set the slope's sign.

Carried through the rung sizes by the class means' variance fraction, $c$ follows the ladder's class-mean part across the eight recordings ($r = 0.99$, about $0.04$ too negative because the full set's $W$ is held fixed), and with the mean-field within-class part the sum gives the shift's sign in eight of eight (mean absolute error $0.038$). Computed on one temporal half and compared with the other, it ranks the recordings at $r = 0.86$ and overshoots, since the mean field has no sample-size term.

\paragraph{The calibrated model.} A generative model calibrated to GT3 draws class means with the fitted multipole correlations and within-class variability as low-rank modes \citep{huang2019} of rank set by the measured within-class PR ($\approx 20$). Modes shared across classes give $+0.080 \pm 0.002$ (five seeds), a third of GT3's shift: they add no directions as classes accumulate. Modes that follow each class's mean as a gain, the multiplicative part of shared cortical variability \citep{lin2015}, give $+0.491 \pm 0.005$, so the two bracket the data, and isotropic within-class variability of equal variance gives $-0.57$. Each class's fluctuations are then the shared modes reweighted by that class's mean, so two classes overlap as the square of their class-mean correlation (within $0.01$ at full gain), while the drifting-grating recordings' classes overlap 1.1 to 5.2 times that square. Cortex's classes share fluctuation structure beyond what their means share, which the two-term account measures directly.

\paragraph{One estimate, several floor draws.} GT3's direction shift appears as $+0.240$ (Figure~\ref{fig:ladder}), $+0.237$ (Appendix~\ref{app:orient}), $+0.234$ (Appendix~\ref{app:random}), $+0.235$ and $+0.232$ (Appendix~\ref{app:rg}), $+0.233$ (Table~\ref{tab:split}) and $+0.231$ (Appendix~\ref{app:worked}, 20 draws): the same ladder against different floor draws, all within 1.5 standard deviations of the mean over 20 seeds of \texttt{rung}'s run with a 20-draw floor ($0.234 \pm 0.004$; Appendix~\ref{app:worked}).

\begin{table}[h]
\centering
\footnotesize
\setlength{\tabcolsep}{3.5pt}
\begin{tabular}{@{}lccc@{}}
\toprule
Recordings (mice) & Neurons or units & Classes & Samples \\
\midrule
\multicolumn{4}{@{}l}{\emph{Stringer et al., two-photon calcium imaging, mouse V1}} \\
\quad Gratings: drifting 3, localized 3, low-contrast 2 (3) & 11{,}311--22{,}565 & 8 directions & 3{,}278--5{,}170 \\
\quad Static gratings 2 (2); spontaneous 8 (6) & 9{,}392--18{,}982 & $\sim 4^\circ$; none & 4{,}470--5{,}356; 5{,}000 \\
\multicolumn{4}{@{}l}{\emph{Allen Brain Observatory, Neuropixels spike counts, ten visual areas}} \\
\quad 32 sessions (32), 167 area populations & 51--421 & 8 directions & 251--600 \\
\multicolumn{4}{@{}l}{\emph{Smith and Kohn, Utah arrays, anesthetized macaque V1}} \\
\quad 3 arrays (3 monkeys) & 59--105 & 12 directions & 2{,}400 \\
\bottomrule
\end{tabular}
\caption{Datasets \citep{stringer2021,stringer2019science,allen2019,smith2008}. Samples are presentations (frames when spontaneous): 390--681 per class in gratings, typically 75 in Allen (VISp: primary visual cortex), 200 in macaque. GT1--GT3: grating mice.}
\label{tab:data}
\end{table}

\paragraph{Data and code.} All analyses use
public data
\citep{stringer2021,stringer2019science,allen2019,kohn2016crcns}; all random
seeds are fixed, and code and data to reproduce every reported number
are at \url{https://github.com/adilamin89/structure-in-the-zoom}.

\section{Wishart derivation of the floor formula}
\label{app:wishart}
For a centered Wishart matrix $S = \nu^{-1}X^\top\!X$ with
$\nu = n{-}1$ degrees of freedom and population covariance
$C = \operatorname{diag}(\lambda_1,\dots,\lambda_d)$, the first two
trace moments are
\begin{align}
  \mathbb{E}[\operatorname{Tr} S] &= \textstyle\sum\lambda_i,
  \label{eq:tr1}\\
  \mathbb{E}[\operatorname{Tr} S^2] &= \tfrac{n}{n{-}1}
    \textstyle\sum\lambda_i^2
    + \tfrac{1}{n{-}1}\bigl(\textstyle\sum\lambda_i\bigr)^{\!2}.
  \label{eq:tr2}
\end{align}
Their ratio gives Eq.~\ref{eq:null}:
\[
  \frac{\bigl(\mathbb{E}[\operatorname{Tr}S]\bigr)^2}
       {\mathbb{E}[\operatorname{Tr}S^2]}
  = \frac{d_{\mathrm{eff}}\,(n{-}1)}{d_{\mathrm{eff}}+n},
  \qquad
  d_{\mathrm{eff}} = \frac{(\sum\lambda_i)^2}{\sum\lambda_i^2},
\]
so the ratio-of-expectations PR depends on the spectrum only through
$d_{\mathrm{eff}}$. PR measures variance concentration; counting latent components is a different estimation problem \citep{vaccari2026}.

For the homogeneous population of
\citet{mazzucato2016} ($N$ units with equal variances $\sigma^2$ and
equal pairwise correlations $\rho$), the denominator
$\mathbb{E}[\operatorname{Tr}S^2]$ coincides with that of their
Eq.~16, and their numerator exceeds ours by $2N\sigma^4/(n-1)$, a term
of order $1/n$; Eq.~\ref{eq:null} is the general-spectrum form of their
finite-trial result. The same crossover form approximates the PR of a random recurrent network's fluctuations measured over a time much longer than their correlation time, and of its mean responses across a finite number of independent contexts \citep[Eqs.~15 and 24]{zhao2026random}, and the PR of the input currents of $M$ units that each copy one of $N$ independent, equal-variance inputs chosen at random, $N/(1 + N/M)$ for large $M$ and $N$ \citep[Eq.~2 at $K = 1$]{litwinkumar2017}: a finite number of units enters as a finite number of trials does, which is why the neuron-axis ladders have floors too.

\paragraph{Against simulation, and the inversion.} Equation~\ref{eq:null} was validated against
direct simulation of isotropic populations over $n = 50$--$4000$ and
$d = 50$--$1000$: the ratio-of-expectations form $d(n-1)/(d+n)$ has
worst-case error $0.36\%$, and a delta-method correction $<0.01\%$. Mapping every rung of both arms through the isotropic inversion of Eq.~\ref{eq:null}, $\PR \to n\,\PR/(n-1-\PR)$, raises GT3's direction shift from $+0.233$ to $+0.249$: the inversion adds $-\log(1 - \PR/(n-1))$ to each rung, more on the floor's rungs, which sit at higher PR ($45$ against the declared $29$ at the first rung), plus $\log[n/(n-1)]$, which both arms share at matched rung sizes, and the two shifts differ by exactly the slope of the first term's difference between the arms ($+0.016$). On the random ladder both versions are near zero ($+0.011$, $+0.007$). The empirical floor needs no spectral assumption and is used throughout.

\section{Worked example: one recording, one command}
\label{app:worked}
This appendix runs the measurement on one recording with one command and shows how to read what it returns. The measurement is released as \texttt{rung} 1.9.0 \citep{rung2026},
\begin{center}\small\url{https://github.com/adilamin89/structure-in-the-zoom}\end{center}
where \texttt{pip install -e .} in a clone installs it; the command below runs in about 70 minutes on one laptop core, most of it the 200 permutations and a floor for each of the 50 random orders, whose classes differ in size.

On GT3 (4{,}442 presentations by 11{,}311 neurons, eight direction
classes in angular order),
\begin{center}
\footnotesize\texttt{rung data gt3\_X.npy gt3\_labels.npy --sectors --decompose --n-perm 200 --out gt3.json}
\end{center}
returns the quantities of the main text in one file (20 floor draws,
against the paper's 10):
\begin{center}
\small
\begin{tabular}{lcc}
\toprule
Quantity & \texttt{rung} & Paper \\
\midrule
$\tobs$, direction ladder & $0.305$ & $0.305$ (Section~\ref{sec:results}) \\
$\tfloor$ (standard error) & $0.074$ ($0.004$) & $0.065$ (Figure~\ref{fig:ladder}) \\
$\delta$ & $+0.231$ & $+0.240$ (Table~\ref{tab:split}: $+0.233$) \\
label permutations & $z = 13.0$, two-sided $p = 1/201$ & $p = 1/201$ \\
within-class part ($D$) & $+0.358$ & $+0.362$ \\
class-mean part ($T_b + C_b + C_x$) & $-0.127$ & $-0.129$ \\
harmonics $a, b_2, c_1, c_3, b_4$ & $0.39, 0.33, 0.14, 0.064, 0.072$ & Section~\ref{sec:symmetry} \\
\bottomrule
\end{tabular}
\end{center}

\paragraph{Matching the paper's numbers.} The two differ by their floor draws: over 20 floor seeds the same command gives $\delta = 0.234 \pm 0.004$ (mean $\pm$ SD), and with the floor drawn as the paper's generators draw it (\texttt{--paper-floor} for the imaging, Allen and macaque generators, which draw ten floor ladders with seed 42 rung by rung; for the lattices \texttt{--floor-by-rung} with their generators' seeds and draw counts in \texttt{--floor-seed} and \texttt{--floor-draws}), \texttt{rung} returns the printed shifts and splits of GT3 (Table~\ref{tab:split}), an Allen population, the three macaque arrays and three lattice ensembles exactly (to float32 rounding where a generator computed in float32).

\texttt{rung}'s $D$ is the within-class part; $T_b + C_b + C_x$ splits the class-mean part into the class means' share of the trace, their own concentration and their overlap with the fluctuations. \texttt{rung}'s permutation $p$ is two-sided (deviation from the null mean in either direction), the paper's one-sided; both reach the minimum $1/201$. Passing the tuned neurons with their
direction-selectivity scores as \texttt{--split-by} runs the equal-size thirds of Section~\ref{sec:symmetry}, and \texttt{--cov-floors} returns the principal-angle dissimilarity of Appendix~\ref{app:allen} with its analytic floor (\texttt{--cov-perm} adds permutation floors to check it).

\paragraph{Language models.} A language model is read at every layer from one command. With the topic axis of Appendix~\ref{app:llm} shipped in the repository (eight topics, 32 prompts each, as JSON),
\begin{center}
\footnotesize\texttt{rung llm --model EleutherAI/pythia-160m --axis axes/world\_knowledge32.json --paper-seeds --match-last-token --out lm.json}
\end{center}
runs in about 19 minutes on one laptop core, model encoding included, and returns, at every layer, the declared and the order-averaged shift with their permutation nulls, the same shift with units standardized, and the last-token report:
\begin{center}
\small
\begin{tabular}{lccc}
\toprule
Layer & order-averaged shift ($p$) & standardized & final-token matched / size-matched \\
\midrule
embedding & $+0.034$ ($0.002$) & $+0.027$ & $-0.001$ / $+0.025$ \\
6 of 12 & $+0.058$ ($0.002$) & $+0.050$ & $+0.033$ / $+0.044$ \\
12 of 12 & $+0.073$ ($0.002$) & $+0.070$ & $+0.046$ / $+0.060$ \\
\bottomrule
\end{tabular}
\end{center}

The declared, order-averaged and standardized shifts equal those of Appendix~\ref{app:llm} to within $2 \times 10^{-9}$ (final-token matching draws its own subsets); the linear and standardized shifts agree in sign at all 13 layers; the classes end on 25 to 31 distinct final tokens. \texttt{rung}'s $p$ is two-sided at its resolution (500 permutations, $1/501$).

\paragraph{What else \texttt{rung} reads.} The tool can also be used on other data: each option reads one quantity on any samples-by-features array with one label per row, or on every layer of a language model, and the last column says where this paper uses it:
\begin{center}
\footnotesize
\begin{tabular}{>{\raggedright\arraybackslash}p{0.30\linewidth}>{\raggedright\arraybackslash}p{0.42\linewidth}>{\raggedright\arraybackslash}p{0.18\linewidth}}
\toprule
Command or option & What it returns & In the paper \\
\midrule
\texttt{rung data X labels} & the floor and its standard error, the shift and the order-averaged shift against label permutations; ladder figure (\texttt{--plot}) & Sections~\ref{sec:decomp}, \ref{sec:results} \\
\texttt{--decompose} (\texttt{--n-bias}) & the within-class and class-mean parts, the sign law of Eq.~\ref{eq:sign}, the finite-sample bias null & Sections~\ref{sec:sign},~\ref{sec:account}, App.~\ref{app:methods} \\
\texttt{--account} & the within-class part beside the overlap law and the identity at the rungs' sample counts, the class-mean part beside Eq.~\ref{eq:sign} carried through the rungs, their sum beside the shift & Section~\ref{sec:account}, Figure~\ref{fig:twoterm} \\
\texttt{--sectors} & the harmonics of the class-mean kernel on a cyclic axis & Section~\ref{sec:symmetry} \\
\texttt{--split-by} & equal-size feature subsets by a score, each against its own floor & Section~\ref{sec:symmetry} \\
\texttt{rung features} (\texttt{--positions}) & a ladder over features, such as the neighbours of anchor cells, against random features & Section~\ref{sec:results}, App.~\ref{app:neuron} \\
\texttt{--block-by}, \texttt{--block-size} & the shift on blocks of features pooled by a score or at random, each against its own floor & App.~\ref{app:rg} \\
\texttt{--order}, \texttt{--antipode} & a declared accumulation order; with antipodal pairs, the share of the climb left after half the circle & App.~\ref{app:order}, \ref{app:methods} \\
\texttt{--candidates} (\texttt{--candidate-strata}, \texttt{--halves}) & each candidate label's shift against its permutations, optionally within the other labels' cells and with discovery on one half and replication on the other & Section~\ref{sec:intro}, App.~\ref{app:allen} \\
\texttt{--strata}, \texttt{--time-blocks}, \texttt{--null-shift} & nulls that keep a nuisance, an acquisition block or the labels' time order fixed & App.~\ref{app:allen}, \ref{app:llm}, \ref{app:methods} \\
\texttt{--valid}, or \texttt{--sample-times} with \texttt{--invalid-intervals} & a data release's validity mask before every statistic & Section~\ref{sec:allen} \\
\texttt{--cov-floors} & the principal-angle dissimilarity and its label-blind floor & Section~\ref{sec:discussion}, App.~\ref{app:allen} \\
\texttt{--paper-floor} (\texttt{--floor-by-rung}, \texttt{--floor-seed}, \texttt{--floor-draws}) & the floor drawn as the paper's generators draw it, which replays their printed values & App.~\ref{app:worked} \\
\texttt{rung llm} (\texttt{--revision}, \texttt{--match-last-token}) & every layer's shift in both feature frames, at any training checkpoint, with the final-token control & App.~\ref{app:llm} \\
\texttt{rung llm --heads} & every attention head's write along the axis, scored by its permutation $z$, beside its class-mean share & Section~\ref{sec:observable}, App.~\ref{app:llm} \\
\texttt{rung axis}, \texttt{rung summarize}, \texttt{rung plot} & a prompt axis from a Hugging Face dataset; a plain-language reading; the depth profile & App.~\ref{app:llm} \\
\texttt{rung nwb} & trials-by-units counts, labels and time windows from an NWB~2 file & Section~\ref{sec:allen} \\
\texttt{rung report} & one self-contained HTML page per result: the figure, the reading, the numbers with their settings & App.~\ref{app:worked} \\
\bottomrule
\end{tabular}
\end{center}

On an ordered axis the declared order is one path among the classes' orders: \texttt{rung} prints the shift's standard deviation over random orders, and where it is not small against the shift, the order-averaged shift is the reading (Appendix~\ref{app:orient}). The sampling threshold of Section~\ref{sec:decomp} compares each class's sample count with that class's own fluctuation dimension (Appendix~\ref{app:random}); the PR($W$) that \texttt{--decompose} prints pools the classes and differs from it.

\end{document}